\documentclass[11pt, epsfig]{article}
\usepackage{epsfig, amsmath, amssymb, amsthm, times}
\usepackage{graphicx}
\newtheorem {thm}{Theorem}[section]
\newtheorem {lem}[thm]{Lemma}

\theoremstyle{defintion}
\newtheorem {df}[thm]{Definition}

\theoremstyle{remark}
\newtheorem{rem}[thm]{Remark}
\theoremstyle{example}

\theoremstyle{assumption}

\def\pf{{\it Proof.\;}}

\def\E{{\mathbb E}}
\def\P{{\mathbb P}}
\def\R{{\mathbb R}}
\def\N{{\mathbb N}}

\def\Z{{\mathbb Z}}
\def\SS{{\mathbb S}}

\def\t{\intercal}

\def\lbl{\label}
\def\be{\begin{equation}}
\def\ee{\end{equation}}
\def\p{\partial}
\def\qed{\square}

\DeclareMathOperator{\re}{Re}

\title{Small-world networks of {K}uramoto oscillators
}
\author{Georgi S. Medvedev 
\thanks{
Department of Mathematics, Drexel University, 3141 Chestnut Street,
Philadelphia, PA 19104, {\tt medvedev@drexel.edu} 
}
}

\begin{document}
\maketitle
\begin{abstract}
The Kuramoto model of coupled phase oscillators on small-world (SW)
graphs is analyzed in this work.
When the number of oscillators in the network goes to infinity, the model acquires 
a family of steady state solutions of degree $q$, called $q$-twisted states.
We show that this class of solutions plays an important role in the formation
of spatial patterns in the Kuramoto model on SW graphs. In particular, the 
analysis of $q$-twisted states elucidates the role of long-range random connections
in shaping the attractors in this model.

\vspace{1 mm}
We develop two complementary approaches for studying  $q$-twisted states
in the coupled oscillator model on SW graphs:  linear stability analysis and numerical
continuation. The former approach shows that long-range random connections 
in the SW graphs promote synchronization and yields the estimate of the synchronization
rate as a function of the SW randomization parameter. The continuation shows
that the increase of the  long-range connections results in patterns
consisting of one or several plateaus separated by sharp interfaces. 

\vspace{1 mm}
These results elucidate the pattern formation mechanisms in 
nonlocally coupled dynamical systems on
random graphs.
\end{abstract}

\section{Introduction}\lbl{sec.intro}
\setcounter{equation}{0}
Coupled dynamical systems on graphs model the evolution of complex
networks of diverse nature. Examples include neuronal 
networks in biology \cite{BulSpo09}; Josephson junctions and 
coupled lasers in physics 
\cite{LiErn92, WatStr94};
communication, sensory, and power networks in technology \cite{DorBul12}, 
to name a few.
In these models, each node of the graph supports a local dynamical system,
whose dynamics is  typically well understood. Local systems at the adjacent nodes
interact with each other, thus, producing collective dynamics of the coupled system.
The spatial structure of the network may vary from regular as in lattice dynamical systems
to  completely random. Thus, it is not surprising that
dynamical systems on graphs exhibit a wide range of spatio-temporal dynamics.

The main challenge in studies of dynamical networks has been relating the structure 
of the graph of the network to its dynamics. In view of the enormous 
complexity of this class of systems, our understanding of the link between the structure
and dynamics has been guided by the analysis of certain successful models illuminating 
various aspects of network dynamics. In this respect the Kuramoto model of coupled oscillators
stands out as one of the most influential in the field \cite{Kur84}. Derived as the phase description of 
large ensembles of limit cycle oscillators, the Kuramoto model has been used by physicists and biologists
to gain insights into many important effects in large networks such as synchronization,
 multistability, phase transitions, wave propagation, and
spatio-temporal chaos, to name a few \cite{Kur84-book}.
Recent experiments
with the Kuramoto model with nonlocal coupling led to the discovery of chimera states, spectacular
patterns combining coherent and irregular turbulence-like behavior \cite{KurBat02}.

\begin{figure}
\begin{center}
{\bf a}\;\includegraphics[height=1.7in,width=1.9in]{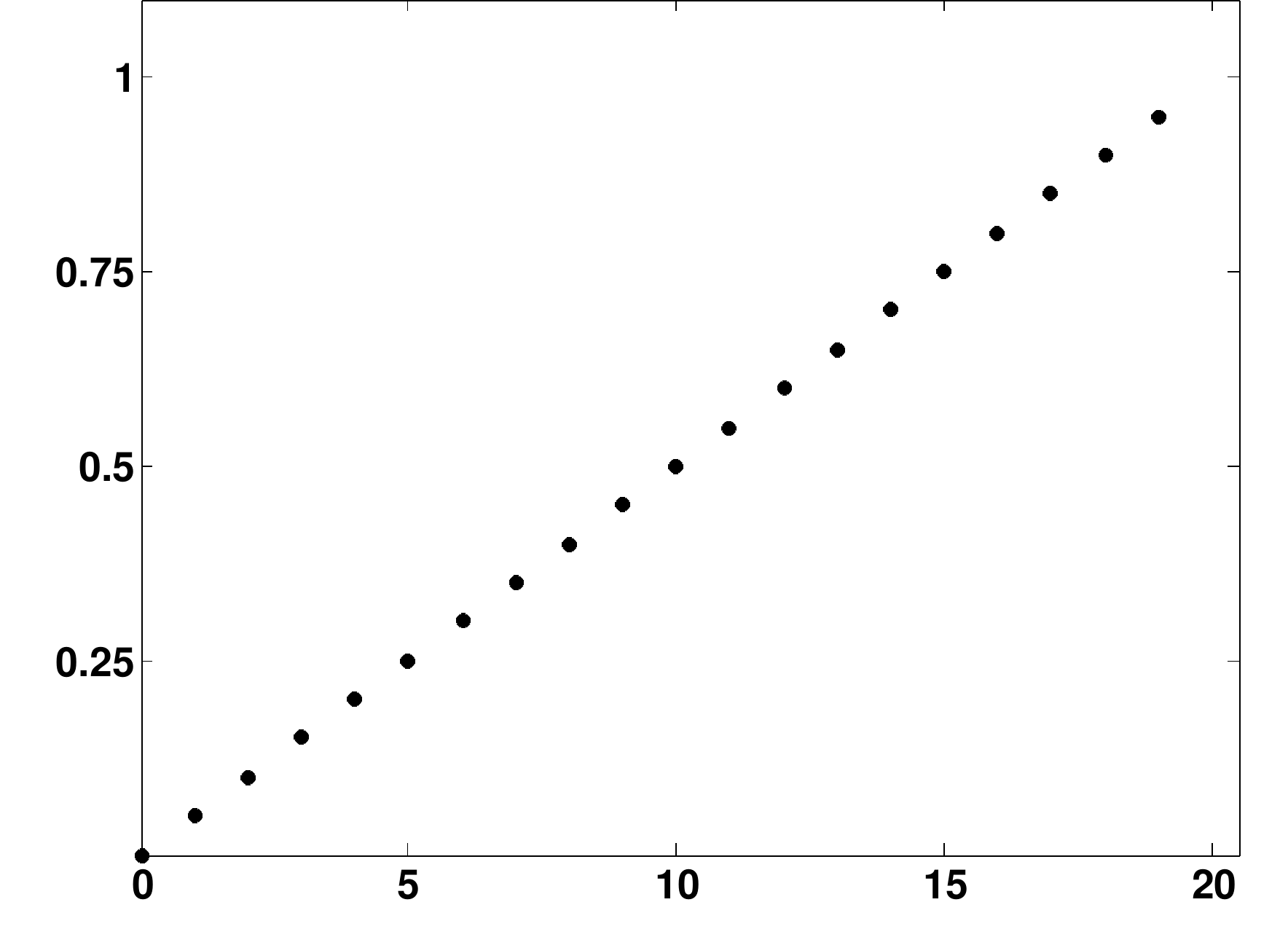}\qquad
{\bf b}\;\includegraphics[height=1.7in,width=1.9in]{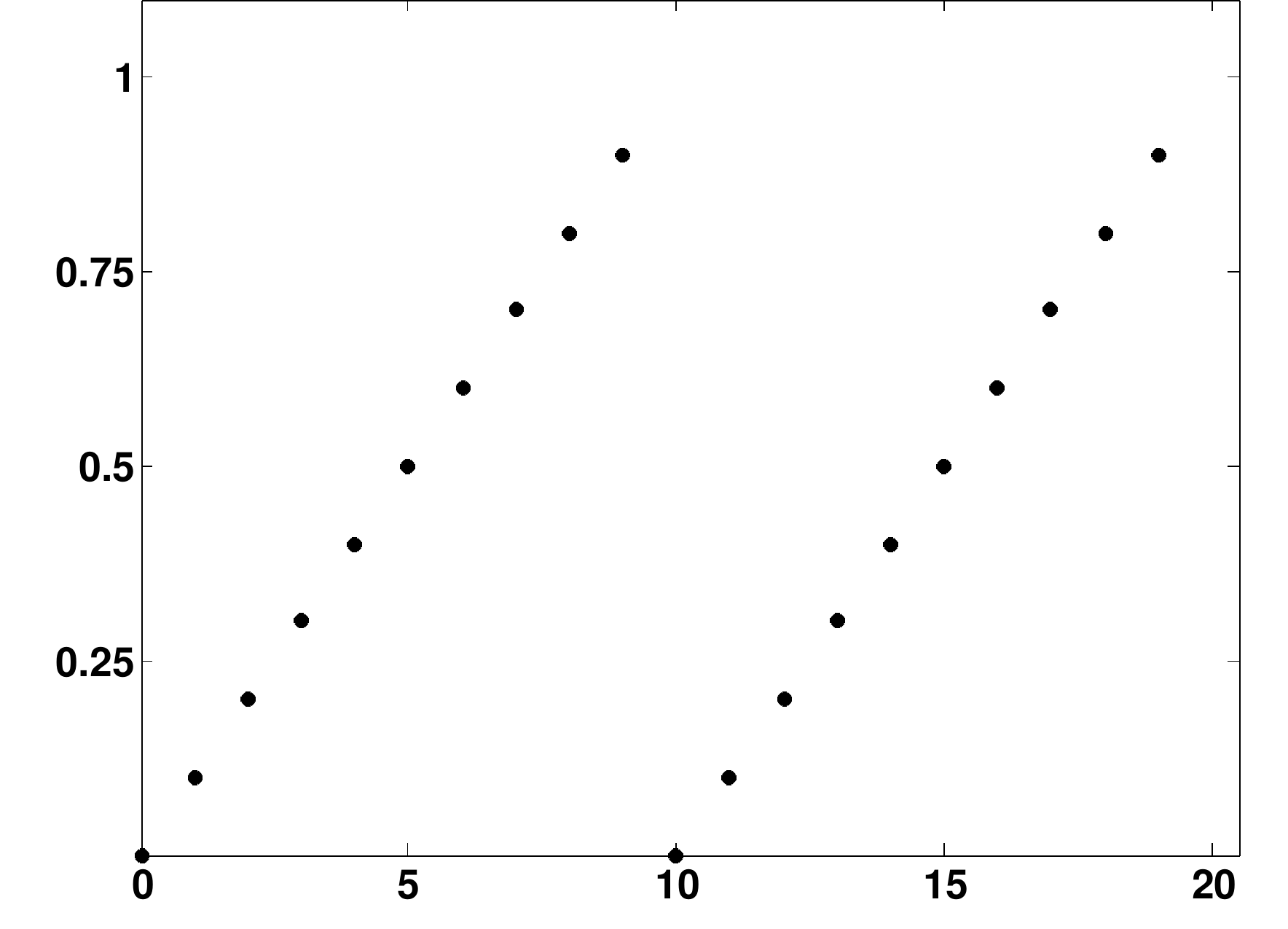}
\end{center}
\caption{Plots of $1$- and $2$-twisted states, steady state solutions
of the Kuramoto model of $20$ coupled oscillators, are shown in
\textbf{a} and \textbf{b} respectively.
}
\lbl{f.-1}
\end{figure} 

The Kuramoto model describes dynamics of coupled phase oscillators on a graph. 
Thus, before we introduce the model, we need to review some graph-theoretic
notation. Let $G=\langle V(G), E(G)\rangle$ be an
undirected graph,  where $V(G)$
stands for the set of nodes of $G$, and the edge set $E(G)\subset V(G)^2$
contains the unordered pairs of connected nodes.  Furthermore, $G$ is called
a simple graph, if it does not contain loops or multiple edges. 

Let $\{G_n, n\in \N\}$ be a sequence of simple graphs. If $|E(G_n)|=O(|V(G_n)|^2)$ then
graphs $G_n, n\gg 1,$ are called dense. 

The following  graph sequence will appear frequently throughout
this paper.  For sufficiently large $n\in\N$ and $k\in\N$ such that $2k<n$, we define
a sequence of $k$-nearest-neighbor ($k$-NN) graphs $\{C_{n,k}\}$ as follows
$$ 
V(C_{n,k})=[n]\quad \mbox{and}\quad E(C_{n,k})=\{ (i,j)\in [n]^2:~ 0<d_n(i,j)\le k\},
$$
where $d_n(i,j) = \min\{|i-j|, n-|i-j|\}$. Let $r\in (0,1/2)$ then $\{C_{n,\lfloor rn\rfloor}\}$ is a
dense graph sequence.

We now introduce the Kuramoto model of coupled  phase oscillators on $\{G_n\}$
\be\lbl{preKuramoto}
{d\over dt} u_{ni}(t) =\omega_i +{\sigma\over n}\sum_{j:~(i,j)\in
  E(G_n)} \sin (2\pi (u_{nj}(t)-u_{ni}(t))), \; i\in [n],
\ee
where $u_{ni}:\R^+\to \SS^1:=\R/\Z,  i\in [n],$ models the evolution of the phase oscillator~$i$, and
 $\omega_i$ is its intrinsic frequency. The scaled sum on the right hand side of (\ref{Kuramoto})
describes the interactions between oscillator~$i$ and the oscillators connected with it.
In the classical Kuramoto model, $\sigma=1$. This is so-called
attractive coupling case, because when $\sigma=1$, synchrony is stable. Recently, 
there has been an interest in studying the Kuramoto model  with $\sigma=-1$, i.e., the repulsive 
coupling case \cite{GirHas12}.
Existence of steady state solutions clearly does not depend on the sign of $\sigma$. However, the stability
of these solutions does. 
In this paper, we are interested in the case of  dense graphs $\{G_n\}$, which corresponds to
the nonlocal coupling in the Kuramoto model (\ref{preKuramoto}).

If all intrinsic frequencies  $\omega_i, i\in [n],$ are equal, as we
assume for the remainder of this paper, they can be eliminated by recasting
(\ref{preKuramoto}) in the rotating frame of coordinates.
Thus,  we consider 
\be\lbl{Kuramoto}
{d\over dt} u_{ni}(t) = {\sigma\over n}\sum_{j:~(i,j)\in
  E(G_n)} \sin (2\pi (u_{nj}(t)-u_{ni}(t))), \; i\in [n].
\ee

Our work was inspired by the analysis of the Kuramoto model on the sequence of $k$-NN graphs
$\{C_{n,k}\},$ $k=\lfloor rn\rfloor, r\in (0,1/2)$ in \cite{WilStr06}. In
this paper, Wiley, Strogatz, and Girvan identified an important class of 
steady state solutions  of degree $q$ $(0\le q\le n)$, 
\be\lbl{q-state}
u^{(q)}_n=(u^{(q)}_{n1},u^{(q)}_{n2},\dots, u^{(q)}_{nn}), \;
u_{ni}^{(q)}={q(i-1)\over n}+c \pmod{1}, \;  c\in [0,1), i\in [n].
\ee
These solutions are called $q$-twisted states.
The $0$-twisted states correspond to spatially uniform (synchronous) solutions. 
Steady states with $q\neq 0$ form linear patterns that wind around $\SS^1$ $q$ times  
(see Fig.~\ref{f.-1}).  It was shown in \cite{WilStr06}  that the stability
of $q$-twisted states for the Kuramoto model depends on the range of nonlocal coupling $r$. 
This provides a nice case study illustrating how the network topology is shaping the structure of the
phase space of the coupled system. 
The follow-up study of  $q$-twisted states in the Kuramoto model with repulsive coupling
suggests their relevance to the chimera states and, more generally,
to the coherence-incoherence transition in coupled systems \cite{GirHas12}.

Interestingly, the original motivation for the analysis in \cite{WilStr06} was the dynamics on 
SW networks. SW graphs, proposed by Watts and Strogatz,  interpolate between graphs
with regular local connections and completely random
Erd\H{o}s-R\'{e}nyi graphs \cite{WatStr98}.
In the classical Watts-Strogatz
SW graph, one starts with the $k$-NN graph on a circle $C_{n,k}$ ($n>2k$). Then each edge
in $C_{n,k}$ is replaced with probability $p\in [0,1]$ to a randomly
chosen one (see Fig.~\ref{f.-2}). We will refer
to $p$ as the randomization parameter. If $p=1$ then $C_{n,k}$ is transformed into the 
Erd\H{o}s-R\'{e}nyi graph. For analytical convenience
Newman and Watts and independently Monasson used another variant of SW connectivity, 
where the new random edges are added to the graph, 
while no edges are removed \cite{NewWat99, Mon99}.

Like many other random graphs, SW graphs have small average path length, but they also have 
large clustering coefficients inherited from the regular graphs. The combination of these two 
properties has been shown in many real-life networks, including
neuronal networks 
\cite{WatStr98,BasBul06},  the power grid 
of the western United States, and professional networks
\cite{WatStr98}. 
Thus, studying dynamical
systems on SW graphs has the potential of elucidating principles governing dynamics in  
diverse realistic networks. A specific question posed in \cite{WilStr06} is how adding a small number 
of random long range connections to otherwise regular network affects stability of the synchronous state.
The synchronizing effect of long-range connections was suggested by the previous studies of related models
\cite{NieSch91, BarPec02}. However, the rigorous understanding of this effect in SW networks was lacking.

In this paper, we study the role of the long-range random  connections in shaping stationary solutions
of the Kuramoto model on SW graphs. For this, we use two complementary approaches: the linear 
stability analysis of $q$-twisted states and numerical continuation. Neither of this approaches can
be applied to the problem as it stands, and, therefore, some preliminary work is required. We comment
on the stability analysis first. In \cite{WilStr06}, Wiley, Strogatz, and Girvan obtained a sufficient condition
for stability of $q$-twisted states for  the $k$-NN coupled Kuramoto model.
Their analysis employed the continuum limit of the Kuramoto model. Unlike the
regular $k$-NN coupling case, on SW graph when the number of oscillators $n$ 
is finite, $q$-twisted states are no longer solutions of the Kuramoto equation. We show that they
become  steady state solutions again in the appropriate probabilistic sense in the limit as $n\to\infty$.  
Thus, as before the continuum limit provides a natural setting for the stability analysis. Our previous
work \cite{Med13b} shows how to derive continuum limits for the Kuramoto model on W-random graphs
including the SW graphs. This allows
the stability analysis of the $q$-twisted states for the SW model to be performed in analogy to the analysis
of the deterministic $k$-NN networks in \cite{WilStr06}.

The stability analysis of $q$-twisted states in the continuous setting  clearly shows the 
respective contributions of the regular local and random long-range connections of the SW network
topology to the spectrum of the linearized problem. It shows that  adding new
long-range random connections enhances stability of the synchronous regime, while inhibiting other
$q$-twisted states. Furthermore, the estimate of the leading eigenvalue of the linearized problem
yields the synchronization rate as a function of the randomization parameter $p$.
The synchronization rate grows monotonically with $p$. Thus, our quantitative estimates
of the eigenvalues of the linearized problem show that adding new random connections
in the SW topology strengthens the synchrony.

\begin{figure}
{\bf a}\includegraphics[height=1.8in,width=2.0in]{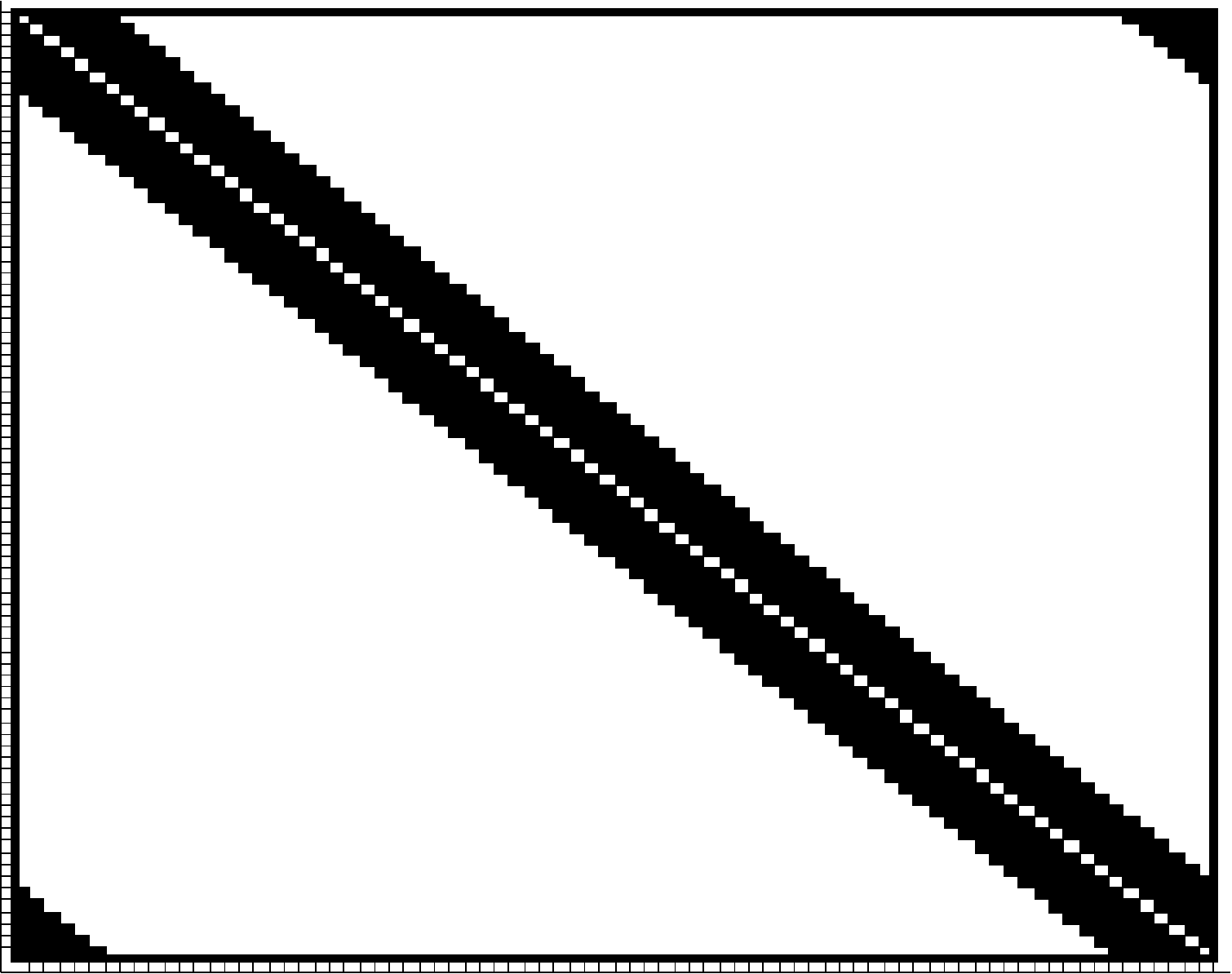}
{\bf b}\includegraphics[height=1.8in,width=2.0in]{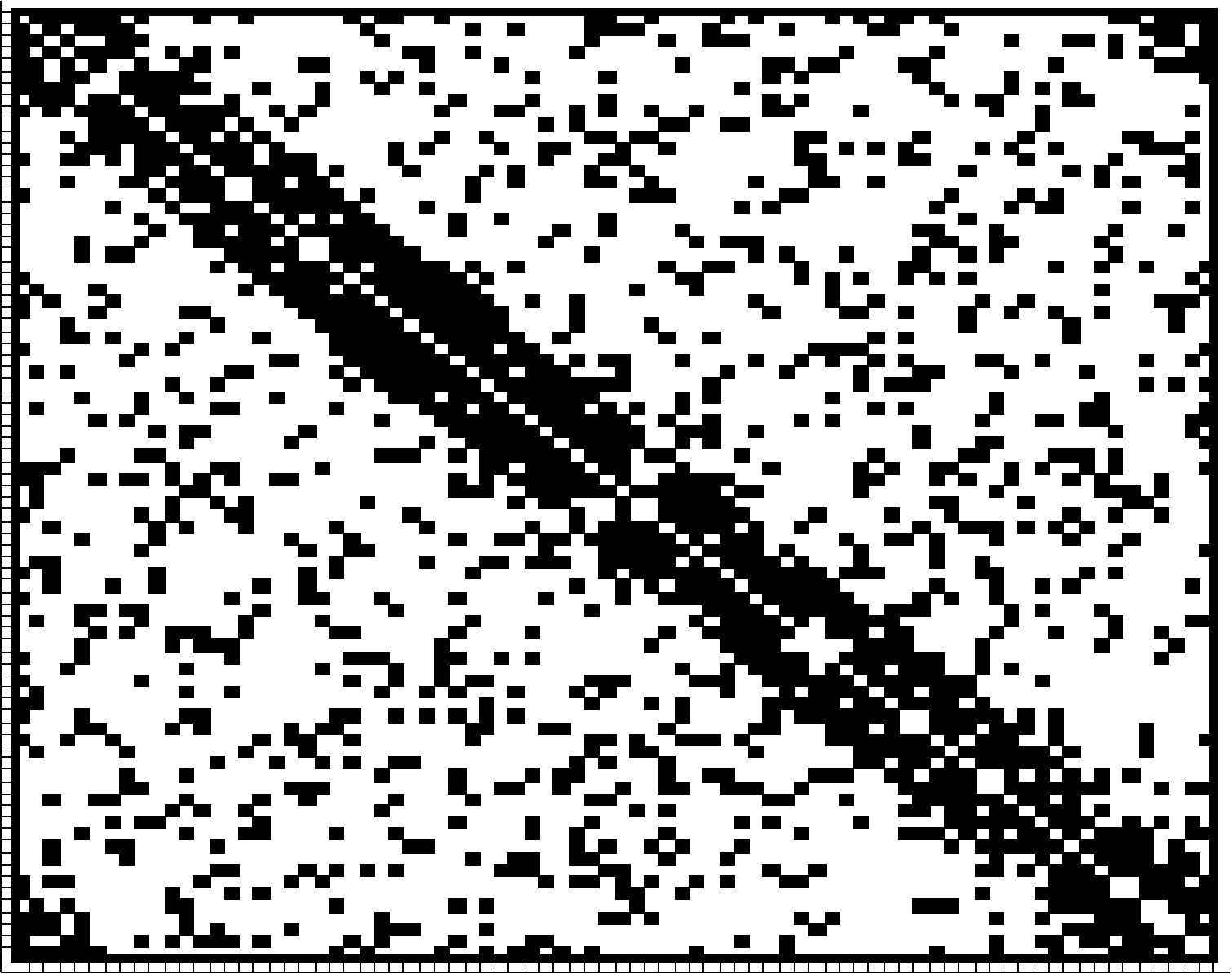}
{\bf c}\includegraphics[height=1.8in,width=2.0in]{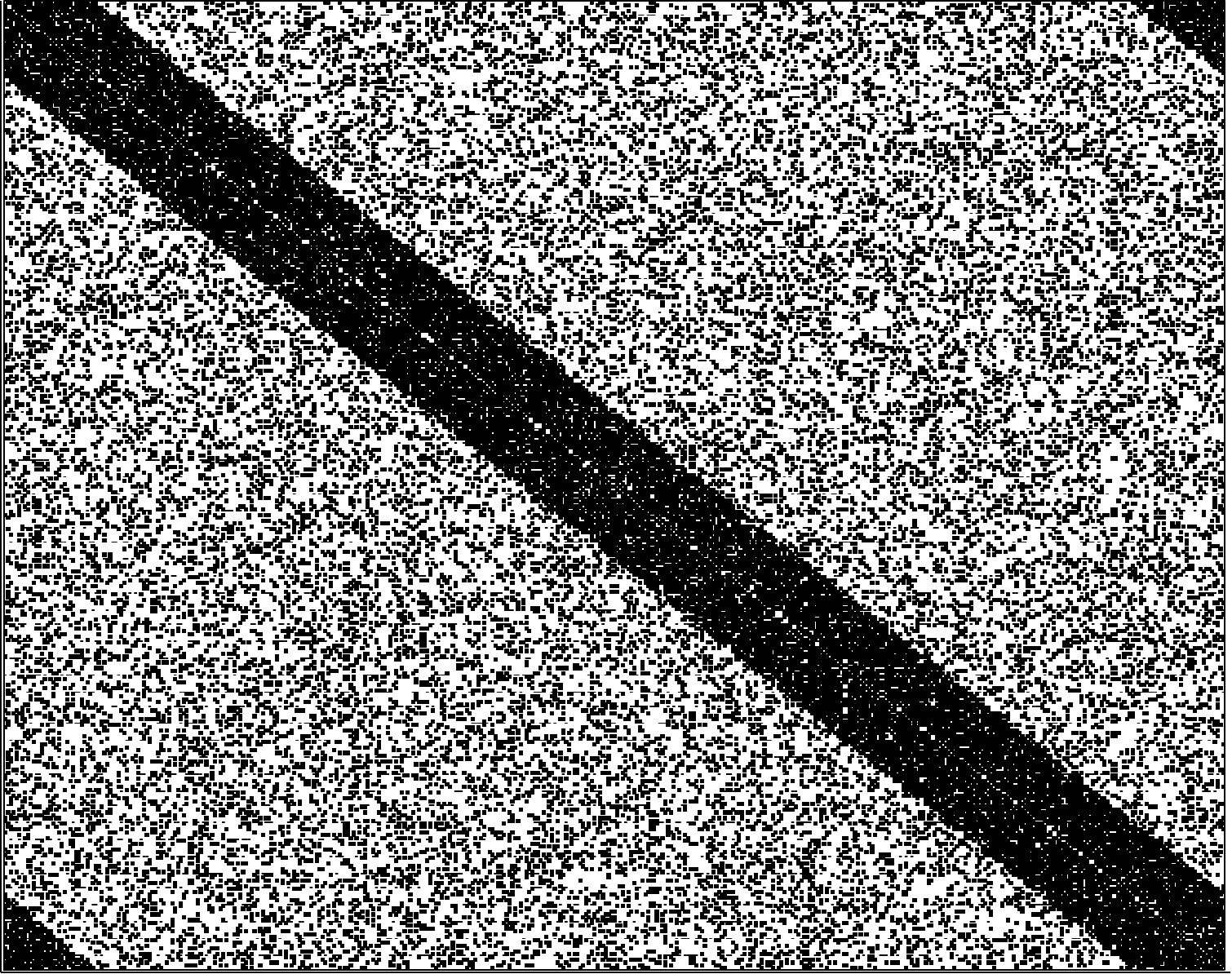}
\caption{\textbf{a}) A graphical representation of the adjacency
  matrix of $C_{n,k}$. In this example $k=6$ and $n=80$, i.e.,
each of the $80$ nodes on the ring is connected to $6$ nearest
neighbors from each side. The horizontal and vertical sides of 
each square are divided into intervals corresponding to the nodes 
of the graph. A black cell at the intersection of a given row and a 
column indicates a pair of connected nodes.
\textbf{b}) The  SW graph obtained by 
replacing each (local) edge in $C_{80,6}$ by a random long-range
connection with probability $p=0.2$.
For increasing values of $p$, the distribution of the black cells 
approaches the uniform distribution, i.e.,
the SW graphs become progressively more similar to the
Erd\H{o}s-R\'{e}nyi graph.
\textbf{c}) The  SW graph obtained by 
replacing each edge in $C_{400,30}$ by a random long-range
connection with probability $p=0.2$.
}
\lbl{f.-2}
\end{figure}

As a complementary approach aimed at elucidating the impact of long-range connections 
on the solutions of the Kuramoto
model and on the spatial patterns that it forms, we developed a continuation method that reveals the 
transformations of the $q$-twisted states under the variation of the randomization parameter $p$. 
The major obstacle in implementing the continuation procedure was the fact that in general the 
right hand side of the Kuramoto model is not continuous in $p$. This is easy to understand
by noting that the adjacency matrices of two independent copies of SW graphs, $G_{n,p_1}$ and $G_{n,p_2}$,
do not have to be close when $|p_2-p_1|$ is small or even when it is
equal to $0$. To overcome this problem we constructed
a special family of SW graphs $G_{n,p}$, whose  members  depend continuously on $p$.
Here, the continuity is meant in the appropriate probabilistic sense
(see Section~\ref{sec.continue} for details). 
We call such family of SW graphs consistent. With the consistent family of SW graphs at hand,
we set up a  numerical continuation algorithm based on the Newton's method and computed
branches of steady state solutions of the Kuramoto model on SW graphs  for $p\ge 0$ by starting from
$q$-twisted states for $p=0$. The numerical continuation revealed an unexpected effect. 
It turns out that for increasing values of $p$, the $q$-twisted states are transformed into
patterns composed of $q$ plateaus separated by interfaces.
We observed this effect in all random realizations of the consistent families of SW 
graphs that we used in our experiments. The formation of plateaus in solutions continued 
from $q$-twisted states is another manifestation of the synchronizing effect  of the long-range 
connections.

This paper is organized as follows. 
In  Section~\ref{sec.cont}, we redefine SW graphs using the framework of
W-random graphs \cite{Med13b} and derive the continuum limit of the Kuramoto
model on SW graphs. In Section~\ref{sec.lin},  we  show that in the limit as $n\to\infty$,
$q$-twisted states become stationary solutions of the Kuramoto model on SW graphs.
This is followed by the linear stability analysis of  $q$-twisted
states. The implications of the stability analysis for synchronization are 
discussed in Section~\ref{sec.sync}. The numerical continuation from the 
$q$-twisted states is developed in Section~\ref{sec.continue}. The final
Section~\ref{sec.discuss} contains concluding remarks.

\section{The continuum limit of the Kuramoto model}
\lbl{sec.cont} \setcounter{equation}{0}
The stability analysis of $q$-twisted states in Section~\ref{sec.lin}
uses the continuum (thermodynamic) limit of the Kuramoto equation (\ref{Kuramoto})
as $n\to\infty$. The thermodynamic limit has proved to be a useful tool for
 elucidating various aspects of the dynamics of large networks 
(see, e.g., \cite{KurBat02, AbrStr06, WilStr06, GirHas12,OmeWol12}).
In \cite{Med13}, we
 provided a rigorous justification of the continuum limit for the models of Kuramoto type
on dense graphs, using the ideas from the theory of graph limits 
\cite{LovGraphLim12,LovSze06,BorChay08}. This work was further extended 
to cover dynamical systems on random graphs in \cite{Med13b}. In particular, the
analysis in \cite{Med13b} shows how to set up and to justify the continuum limit
for the Kuramoto model on the SW graphs.
Before we apply the continuum limit to study the stability of $q$-twisted states,
we  review a few facts illuminating the basis for the 
limiting dynamics of (\ref{Kuramoto}) as $n\to\infty$.

As the first step toward the continuum limit, we  redefine  SW graphs as  W-random 
graphs \cite{LovGraphLim12,LovSze06}. The latter provide a flexible framework for constructing 
random graphs, which is especially
convenient for studying coupled dynamical systems. Below, we review 
the construction of W-random graphs and use it to define SW graphs (cf.~\cite{Med13b}).
 
Let $\mathcal{W}_0$ be a class of symmetric measurable functions on $I^2$ with values in 
$I$. Here and throughout this paper, $I$ stands for the unit interval $[0,1]$. 

\begin{df}\lbl{df.Wrandom}\cite{LovSze06}
Let $W\in\mathcal{W}_0$ and 
$$
X_n=\{x_1, x_2, x_3, \dots, x_n\}
$$ 
be a set of distinct points from $I$. 
Define $G_n=\langle [n], E(G_n)\rangle$ such that 
for each $(i,j)\in [n]^2$
$$
\P\{(i,j)\in E(G_n)\}=\left\{\begin{array}{ll} W(x_i,x_j),& i\neq j,\\ 0,& \mbox{otherwise},
\end{array}\right.
$$
where $\P\{\cdot\}$ stands for the probability of an event.
Graph $G_n$ is called a $W$-random graph and denoted by $G_n=\mathbb{G}(W,X_n)$.
\end{df}

In the remainder of this paper, we use
\be\lbl{def-Xn}
X_n=\left\{0, {1\over n}, {2\over n},\dots, {n-1\over n}\right\}
\ee
and
\be\lbl{def-W}
W(x,y)=\left\{\begin{array}{ll} 1,& d(x,y)\le r,\\ 0, &\mbox{otherwise}, \end{array}\right.
\ee
where $r\in (0,1/2)$ is a fixed parameter, and 
\be\lbl{def-d}
d(x,y)=\min\{ |x-y|, 1-|x-y|\}.
\ee
With these choices of $W$ and $X_n$, $G_n=\mathbb{G}(W,X_n)$ is the $\lfloor rn\rfloor$-NN
graph, which is used as a starting point in the construction of SW graphs
\cite{WatStr98, NewWat99}.

We are now in a position to interpret SW networks as W-random
graphs (cf.~\cite{Med13b}). To this end, let
\be\lbl{GM}
G_{n,p}=\mathbb{G}(W_p, X_n),\;\;\mbox{where}\;\; W_p = (1-p) W + p(1-W), \;p\in [0,0.5].
\ee
Let us call $(i,j)\in [n]^2, i\neq j,$ a local edge
if $(i,j)\in E(C_{n,\lfloor rn\rfloor})$, and a long-range one otherwise.
The construction of $G_{n,p}$ can be interpreted as follows:  $G_{n,p}$
is obtained from $C_{n,\lfloor rn\rfloor}$ by letting the latter to shed each local connection
with probability $p$ and to add each long-range one with the same probability  
(see Fig.~\ref{f.-2}b).  In particular, $G_{n,0}$ coincides with
$C_{n,\lfloor rn\rfloor}$, and $G_{n,0.5}$ is the Erd\H{o}s-R\'{e}nyi
graph with the edge density $1/2$. The construction 
of $G_{n,p}$ resembles that of the Watts-Strogats SW graphs \cite{WatStr98},
but differs from it in some details. For other $W$-random graphs, which match 
the classical  Watts-Strogatz and Manasson-Newman-Watts models of SW graphs 
\cite{WatStr98, NewWat99, Mon99} more closely, we refer an interested
reader to \cite[Section~5]{Med13b}. 
In the remainder of this paper, (\ref{GM}) will be used as a model of SW connectivity.

Now we turn to the main point of this section: the derivation of the continuum 
limit for the Kuramoto model on the sequence of SW graphs $\{G_{n,p}\}$.
To explain the intuition behind the limiting procedure, let us rewrite 
(\ref{Kuramoto}) as
\be\lbl{prelim}
{d\over dt} u_{ni} ={\sigma\over n} \sum_{j=1}^n a_{ij}^{(n)} \sin(2\pi(u_j-u_i)),\; i\in [n],
\ee
where $(a_{ij}^{(n)})$ stands for the $n\times n$ adjacency matrix of $G_{n,p}$.
Our goal is to derive the infinite-dimensional analog of (\ref{prelim}) as $n\to\infty$.
To this end, we define a step-function $u_n: I\times \R$  such that
\be\lbl{step-u}
u_n(x,t)=u_{ni}(t),\; x\in [(i-1)n^{-1}, in^{-1}),\; i\in [n],
\ee
where $u_n=(u_{n1}, u_{n2},\dots, u_{nn})$ is a solution of the discrete model (\ref{prelim}).
We expect that as $n\to\infty$, the sequence of step-functions
$u_n(x,t)$ approaches $u(x,t)$ (in the appropriate norm). The latter function is interpreted
as the solution for the continuum model.

Next, we have to decide how to interpret the adjacency matrix $(a_{ij}^{(n)})$ as $n\to\infty$.
The answer to this question is suggested by the theory of graph limits, which for every
sequence of convergent dense graphs, like $\{G_{n,p}\}$, defines the limiting object - 
a symmetric  measurable function on the unit square. The limit
of a convergent graph sequence is called a graphon \cite{LovGraphLim12}.
From the theory of graph limits, we know that
$\{G_{n,p}\}$, as a sequence of $W$-random graphs,
converges to the underlying graphon $W_p$ (cf.~\cite{LovSze06}).
Below, we provide an informal intuitive interpretation of the graph limit 
$W_p$, which is meant to explain its appearance in the continuum limit 
of (\ref{prelim}). A detailed exposition of the theory of graph limits 
can be found in the monograph by Lov{\'a}sz \cite{LovGraphLim12},
or in the following papers \cite{BorChay08, LovSze06, Pikh10}.

To see the meaning of the graph limit $W_p$, it is instructive to recall
the geometric representation of the adjacency matrices of SW graphs
in Fig.~\ref{f.-2}b,c. The matrices depicted in these plots correspond
to $G_{80,0.2}$ and $G_{400, 0.2}$. These graphs belong to the same sequence
of SW graphs  $\{G_{n,0.2}\}$ generated by 
graphon $W_{0.2}$ with $r=3/40$ (see (\ref{GM})).
Note that for large $n$, as in Fig.~\ref{f.-2}c, black pixels fill different
regions of the unit square  uniformly with two different densities
$1-p$ and $p$, just as in the definition  of the function $W_p$ (see
(\ref{def-W}) and (\ref{GM})). 
In fact, the graphon $W_p$ represents the density of connections
between different subsets of nodes of the SW graph $G_{n,p}$ for large $n$. 

We have identified $u(x,t)$ and $W_p(x,y)$ as continuous counterparts 
of $u_n(t)=(u_{n1}(t), u_{n2}(t),\dots,u_{nn}(t))$ and $(a_{ij}^{(n)})$ in the large $n$ limit.
Furthermore, the right-hand side in (\ref{prelim}) has the form of a Riemann sum.
Therefore, by sending $n\to\infty$ in (\ref{prelim}), we formally arrive at
the following evolution equation
\be\lbl{WKur}
{\p\over\p t} u(x,t) =\sigma \int_I W_p(x,y) \sin\left(2\pi( u(y,t)-u(x,t))\right) dy.
\ee
The continuum limit (\ref{WKur}) was rigorously justified in \cite{Med13b}. 
Specifically, Theorem~3.4 of \cite{Med13b} implies that for any $T>0$, $u_n(x,t)$ (see (\ref{step-u}))
converge to $u(x,t)$ in the $C(0,T; L^2(I))$ norm in probability as $n\to\infty$,
provided the initial conditions $u_n(\cdot,0)$ and $u(\cdot,0)$ are sufficiently close.

\section{The stability analysis of q-twisted states}
\lbl{sec.lin}
\setcounter{equation}{0}
The goal of this section is to elucidate the role of $q$-twisted
states (\ref{q-state}) in the long-time 
asymptotic behavior of solutions of the Kuramoto model (\ref{Kuramoto}) on SW graphs
(\ref{GM}) for large $n$. First, we show that  $q$-twisted states become steady state solutions
of (\ref{Kuramoto}), (\ref{GM}) as $n\to\infty$ with probability $1$ (see Lemma~\ref{lem.LLN}).
Next, we use the continuum limit (\ref{WKur}) to study the linear stability of the continuous
$q$-twisted states
\be\lbl{q-cont}
u^{(q)}(x)=qx+c \pmod{1}, \; c\in [0,1),\; q\in\Z
\ee
(see Theorem~\ref{thm.lin}). The results of this section will be used in Section~\ref{sec.sync}
to study synchronization in SW networks.

Note that while the continuous $q$-twisted states (\ref{q-cont}) solve (\ref{WKur}) for all values of $p$, 
their discrete counterparts generically do not solve
(\ref{Kuramoto}) for $p>0$. They nonetheless remain relevant for the following reasons.
First, for large $n$, (\ref{Kuramoto}) has solutions that remain close
to $u_n^{(q)}$ on finite time intervals (cf.~\cite[Theorem~3.4]{Med13b}). Furthermore, 
the following lemma shows directly (i.e., without invoking
the limiting equation (\ref{WKur})) that $u_n^{(q)}$ become the
steady-state solutions of (\ref{Kuramoto}) for $p>0$ 
in the limit as $n\to\infty$. 

\begin{lem} \lbl{lem.LLN}
Let $\{G_{n,p}\}$ be a sequence of SW graphs (\ref{GM}), (\ref{def-W}) with
$r\in (0,1/2]$ and
\be\lbl{rhs-dKur}
\mathcal{R}_{ni} (u_n)={1\over n}  \sum_{j: (i,j)\in E(G_{n,p})} 
\sin\left(2\pi( u_{nj}-u_{ni})\right),\; i\in [n],
\ee
where $u_n=(u_{n1},u_{n2},\dots, u_{nn})$.
Then for any $q\in \N\bigcup \{0\}$ and $i\in\N$
\be\lbl{conv-rhs}
\lim_{n\to\infty}\;\mathcal{R}_{ni} (u_n^{(q)})=0 
\ee
almost surely.
\end{lem}

\pf\; Assume first that $r\in (0,1/2)$. Let $i\in \N$ be arbitrary but fixed and  let $n\ge i$. To simplify notation, throughout this proof we suppress 
the dependence of various sets and variables on $i$.
Denote 
$$
J_n^{(1)}=\{  j\in [n]:~ j\neq i \;\& \;d(i,j)\le rn \}, \; 
J_n^{(2)}=\{  j\in [n]:~ j\neq i\; \&\; d(i,j) > rn \},  
$$
$n_1:=|J_n^{(1)}|=2\lfloor rn\rfloor$ and $n_2=n-1-n_1.$
Note that for $r\in (0,0.5)$,  $n_{1,2}$ tend to infinity with $n\to\infty.$

Let $\{\xi_{nj}^{(k)}\}_{ j\in J_n^{(1)} }, \, k\in\{1,2\} $ be two  arrays of 
independent identically distributed (IID) random variables (RVs) such that 
\be\lbl{binomial}
\mathcal{ L}(\xi_{n1}^{(1)})=\operatorname{Bin}(1-p)\; \mbox{and}\;
\mathcal{ L}(\xi_{n1}^{(2)})=\operatorname{Bin}(p).
\ee
Here, $\mathcal{L}(\cdot)$ stands for the probability distribution of a RV, and
$\operatorname{Bin}(p)$ denotes the binomial distribution with parameter $p$.
Define
\be\lbl{def-eta}
\eta_{nj}^{(k)}=\xi_{nj}^{(k)} \sin (2\pi(u^{(q)}_{nj}-u^{(q)}_{ni}))=\xi_{nj}^{(k)} \sin \left({2\pi q(j-i)\over n}\right)
, \; j\in J_n^{(k)}, k\in \{1,2\}.
\ee

Using (\ref{rhs-dKur}) and the definitions of $\{\eta_{nj}^{(k)}\}, k\in \{1,2\}$, we have
\be\lbl{rewrite-dKur}
\mathcal{R}_{ni}(u^{(q)})= {n_1\over n}  (n_1^{-1}S^{(1)}_n)
+{n_2\over n}  (n_2^{-1}S^{(2)}_n),
\ee
where
\be\lbl{def-Sn}
S^{(k)}_n= \sum_{j\in J^{(k)}_n} 
\eta_{nj}^{(k)},\; k\in\{1,2\}.
\ee

Using (\ref{binomial}) and (\ref{def-eta}), we estimate 
\begin{eqnarray}
\lbl{2nd}
\E {\eta_{nj}^{(k)}}^2 &=& \sin \left({2\pi q(j-i)\over n}\right)^2 p_k\le 1,\\
\lbl{fourth}
\E {\eta_{nj}^{(k)}}^4 &=& \sin \left({2\pi q(j-i)\over n}\right)^4 p_k\le 1, \; k\in\{1,2\}, j\in [n],
\end{eqnarray}
where $p_1=1-p$ and $p_2=p$.

Further, using (\ref{binomial}) and (\ref{def-Sn}), we have
\be\lbl{est-ESn}
\E S_n^{(k)} = p_{k} \sum_{ j\in J_n^{(k)} } \sin \left({2\pi q(j-i)\over n}\right) =0, \; k\in\{1,2\},
\ee
because $J_n^{(k)}$ is symmetric about $i$.

From this point, the proof follows the lines of the proof of the strong law of large numbers
\cite[Theorem~6.1]{Bill-Prob}. 
Specifically, we estimate
\begin{eqnarray}\nonumber
\E {S_n^{(k)}}^4 &=& \sum_{j_1, j_2,j_3, j_4 \in J_n^{(k)}} 
\E \left[ \eta_{nj_1}^{(k)} \eta_{nj_2}^{(k)} \eta_{nj_2}^{(k)}\eta_{nj_2}^{(k)}\right] \\
\nonumber  
                       &=&  \sum_{j\in J_n^{(k)}} \E \left[{\eta_{nj}^{(k)}}^4\right]+
\dbinom{4}{2}
\sum_{j_1<j_2, j_1,j_2\in J_n^{(k)}} \E \left[{\eta_{nj_1}^{(k)}}^2{\eta_{nj_2}^{(k)}}^2\right] \\
\lbl{S4}
&\le& n+3n(n-1)< 3n^2,\; k\in\{1,2\},
\end{eqnarray}
where we used independence of $\{\eta_{nj}^{(k)}:\, j\in J_n^{(k)}\},$ 
(\ref{2nd}), (\ref{fourth}), and (\ref{est-ESn}).

By Markov inequality, using (\ref{est-ESn}) and (\ref{S4}), for any $\epsilon>0$
we have
\be\lbl{Markov}
\P\{ |S_n^{(k)}|\ge n\epsilon \}\le 
3 \epsilon^{-4} n^{-2},
\ee
which in turn implies via the first Borel-Cantelli lemma \cite[Theorem~4.3]{Bill-Prob}
$$
\P \{ | n_k^{-1} S_n^{(k)} | \ge \epsilon \; \mbox{holds for infinitely many}\; n\}=0, 
\;  k\in\{1,2\}.
$$
By Theorem~5.2(i) in \cite{Bill-Prob}, the last statement is equivalent to convergence of $n_k^{-1}S_n^{(k)}$ to $0$
almost surely. Therefore,  the  combination of (\ref{rewrite-dKur}),  
$$
\lim_{n\to\infty} {n_1\over n}=2 r, \;\mbox{and}\;\lim_{n\to\infty} {n_2\over n}=1-2r,
$$
yields (\ref{conv-rhs}) for $r\in (0,1/2)$.  
If $r=1/2$ then $S^{(2)}_n=0$, and (\ref{conv-rhs}) follows by repeating the above argument
for $S_n^{(1)}$ only.\\ 
$\qed$

Having shown, that $q$-twisted states solve the Kuramoto equation on $\{G_{n,p}\}$ in the 
large $n$ limit, we turn to study the stability of $q$-twisted states, as steady state solutions
of the continuum equation. For concreteness, in the remainder of this paper we restrict to 
the attractive coupling case, i.e., $\sigma=1$ in (\ref{Kuramoto}). The repulsive coupling case
can be treated similarly. Thus, we rewrite (\ref{WKur}) as follows
\be\lbl{cKur}
{\p u(x,t)\over \p t}= \int_{I} K_p(y-x) \sin (2\pi(u(y,t)-u(x,t))) dy,
\ee
where 
\be\lbl{specialize-Kp}
K_p(x)=pG_{1/2}(x)+(1-2p) G_r(x),\quad
G_r(x)=\left\{\begin{array}{cc} 1, & d(x) < r, \\ 0, &\mbox{otherwise}.\end{array}  \right. 
\ee
and $d(\cdot)$ was defined in (\ref{def-d}). 
\begin{rem}\lbl{rem.Kp}
The results of this section hold for a more general class of kernels $K_p$. It is sufficient to
assume that $K_p(\cdot)$ is a even bounded function.
\end{rem}

We are now in a position to formulate a sufficient condition for stability 
of $q$-twisted states $u^{(q)}$ (see (\ref{q-cont})).

\begin{thm}\lbl{thm.lin} 
For $q\in\Z$ and $p\in [0,0.5]$, the $q$-twisted state $u^{(q)}$ is a steady state of
 (\ref{cKur}). Moreover, it is linearly stable with respect to perturbations
 from $L^\infty (I)$ provided
\be\lbl{stab-cond}
\lambda_p(q,m):=\tilde K_p(m+q)-2\tilde K_p(q)+\tilde K(q-m)<0\;\forall m\in\N,
\ee
where
\be\lbl{cos-transform}
\tilde K_p(m)=\int_{I} K_p(x)\cos(2\pi mx),\; m\in \Z.
\ee
\end{thm}

\begin{rem} Theorem~\ref{thm.lin} extends the sufficient condition 
 that was obtained for $k$-NN deterministic networks in \cite{WilStr06}
to networks on SW graphs.
\end{rem}
We precede the proof of Theorem~\ref{thm.lin} with the following observation,
which follows from  the odd-symmetry of the product $K_p(x)\sin(x)$.
\begin{lem}\lbl{lem.zero-mean}
Let $\mathbf{u}\in C^1(\R, L^\infty (I))$ be a solution of (\ref{cKur}). Then
\be\lbl{zero-mean}
{\p \over\p t} \int_{I} u(x,t) dx =0.
\ee
\end{lem}
\pf
By integrating both parts of (\ref{cKur}) over $I$ and using  Fubuni's theorem,
we have
\be\lbl{double-int}
{\p \over\p t} \int_{I} u(x,t) dx = \int_{I\times I} K_p(y-x) 
\sin (2\pi(u(y,t)-u(x,t))) dxdy,
\ee
where we also interchanged differentiation with integration on the left-hand side,
because $\mathbf{u}\in C^1(\R, L^\infty (I))$ and $\mathbf{u}^\prime$ is bounded on $\R$
(cf.~\cite[Theorem~3.1]{Med13}). The integral on the right-hand side
is zero, because  the domain of integration $I\times I$ is symmetric
with respect to the line $y=x$ and the integrand is an odd symmetric function.\\
$\qed$

\pf (Theorem~\ref{thm.lin})
Consider $H=L^\infty(I)$ as a subspace of $L^2(I)$. Then
$$
H=H_0\oplus H_0^\perp,
$$ 
where 
$$
H_0=\left\{ u\in H:~\int_{I} u(x) dx =0\right\}\;\;\mbox{and}\;\;
H_0^\perp=\left\{ u\in H:~u(x)=c \;\mbox{a.e., for some}\; c\in\R  \right\}.
$$
By Lemma~\ref{lem.zero-mean}, $H_0$ and $H_0^\perp$ are invariant under the flow
(\ref{WKur}).  It is easy to see that $q$-twisted state solutions of (\ref{WKur}) remain stable under
spatially homogeneous perturbations from $H_0^\perp$. Therefore, it is sufficient to
study the stability of $q$-twisted states with respect to the perturbations from $H_0$.

To this end, we linearize (\ref{cKur}) about the $q$-twisted state $u^{(q)}$, i.e., we plug 
$ u(x,t)=u^{(q)}(x)+\eta(x,t)$ into (\ref{cKur}) and retain only the linear terms on the
right-hand side of the resultant equation
\be\lbl{linearize}
 {\p \eta(x,t)\over \p t}=2\pi \int_{I} K_p(y-x) \cos(2\pi q (y-x)) \left[\eta(y,t)-\eta(x,t)\right]dy,
\ee
which we supply with the initial condition 
\be\lbl{restrict}
\eta(\cdot, 0)\in H_0.
\ee

We expand $\eta(x,t)$ in the Fourier series
\be\lbl{expand}
\eta(x,t)= \sum_{k\in\Z} \hat\eta_k(t) e^{-i2\pi kx}, \;\mbox{where}\;
\hat\eta_k(t)=\int_{I} \eta(x,t)e^{i2\pi kx}dx.
\ee
Note that $\hat\eta_0\equiv 0$, because $\hat\eta_0(0)=0$ as follows from (\ref{restrict}).

To simplify notation we rescale the time variable in (\ref{linearize}) to absorb the $2\pi$ factor on
the right-hand side. Thus, (\ref{linearize}) can be rewritten
in the following form
\be\lbl{lin}
 {\p \eta \over \p t}=K_p^{(q)}*\eta -(\re \hat K_p(q))\eta,
\ee
where $*$ denotes the convolution and
$$
K_p^{(q)}(x)= K_p(x)\cos(2\pi q x),\;
\hat K_p(m)= \int_{I} K_p(x)e^{i2\pi mx} dx.
$$

By applying the Fourier transform to both sides of (\ref{lin}), we have
\be\lbl{modes}
{d \hat\eta_m\over dt}=\left[ \hat K_p^{(q)}(m)-\re\hat K_p(q) \right] \hat\eta_m,\;
m\in\Z/\{0\}.
\ee

Thus, using Parseval's identity, we arrive at the sufficient condition for linear stability of the $q$-twisted
state
\be\lbl{sufficient}
\re \left[ \hat K_p^{(q)}(m)-\hat K_p(q)\right] <0\;\; \forall m\in\Z/\{0\}.
\ee
Further, it is easy to verify that
$$
\hat K_p^{(q)}={\hat K_p(m+q)+\hat K_p(m-q)\over 2},\; m\in\Z/\{0\}.
$$
Thus, (\ref{sufficient}) can be rewritten as follows
\be\lbl{cond-lambda}
\lambda_p(q,m)=\tilde K_p(m+q)-2\tilde K_p(q)+\tilde K_p(m-q)<0\; \forall m\in\Z/\{0\}, 
\ee
where
$$
\tilde K_p(m)=\re\hat K_p(m)=\int_{I} K_p(x)\cos (2\pi m x)dx, \; m\in\Z.
$$
Since $\tilde K_p(m)$ is an even function, it is sufficient to restrict  (\ref{cond-lambda})
for $m\in \N$.
This concludes the proof of Theorem~\ref{thm.lin}.\\ $\qed$

\section{Synchronization}
\lbl{sec.sync}
\setcounter{equation}{0}
In this section, we apply Theorem~\ref{thm.lin} to study synchronization
in the Kuramoto model on SW graphs (\ref{cKur}).
Throughout this section, we continue to use the rescaled linearized 
equation (\ref{lin}).

Using (\ref{specialize-Kp}) and (\ref{cos-transform}), we have
$$
\tilde K_p(m)=p\tilde G_{1/2}(m)+(1-2p)\tilde G_r(m),
$$
where
\be\lbl{tilde-G}
\tilde G_r(m)=\left\{\begin{array}{cc} 2rS(2\pi mr), & m\neq 0,\\
2r, & m=0, \end{array}\right.
\ee
and $S(x)=x^{-1}\sin x$.

By applying Theorem~\ref{thm.lin} to (\ref{cKur}) with (\ref{specialize-Kp}),
we obtain the following condition for stability of the $q$-twisted states
\be\lbl{stab-q-SW}
\lambda_p(q,m)= pD^2_m \tilde G_{1/2}(q) + (1-2p) D^2_m \tilde G_r(q) <0 \; \forall m\in \N,
\ee
where $D^2_m$ stands for the second-order difference
\be\lbl{2nd-dif}
D^2_m \tilde K(q)=\tilde K_p(q+m)-2\tilde K_p(q)+\tilde K_p(q-m).
\ee

Using (\ref{tilde-G}), 
 we rewrite (\ref{stab-q-SW}) as follows
\be\lbl{other-q}
\lambda_p(q,m)=p(-2\delta_{q0}+\delta_{qm}) + (1-2p)\lambda_0(q,m)<0 \; \forall m\in \N,
\ee
where 
$$
\lambda_0(q,m)= 2r \left[ S(2\pi r(q+m))-2S(2\pi rq)+ S(2\pi r(q-m))\right]
$$
and the Kronecker delta $\delta_{ij}=1$ if $i=j$ and is equal to $0$ otherwise.  

Equation (\ref{other-q}) yields the following sufficient condition for synchronization in
SW networks
\be\lbl{lambda-p}
\lambda_p(0,m)=-2p + (1-2p) \lambda_0(0,m)<0 \; \forall m\in \N,
\ee
where
\be\lbl{lambda-0}
\lambda_0(0,m)= 4r(S(2\pi mr)-1).
\ee

Conditions  (\ref{other-q}) and (\ref{lambda-p})  show how the random
long-range connections in the SW network affect the stability of the $q$-twisted states.
For the $k$-NN coupled networks, the linear stability of the $q$-twisted states
was studied in detail by Wiley, Strogatz, and Girvan \cite{WilStr06}. They found that 
the $q$-twisted state is stable provided $0\le qr <\mu$, where $\mu\approx 0.66$
is a solution of a certain transcendental equation (see \cite [Equation~(18)]{WilStr06}).
For these values of $q$, we have
$$
\lambda_0(q,m)<0\;\forall m\in\N.
$$
Thus, the synchronous solution ($q=0$) is stable for the full range of $r\in (0,0.5)$. In addition,
for small $r>0$, there may be many other stable $q$-twisted states.  
For such values of $r$, these $q$-twisted states remain stable on SW graphs $G_{n,p}$ with 
\be\lbl{qq}
0<p<{-\lambda_0(q,q)\over 1-2\lambda_0(q,q)}.
\ee
From Equation (\ref{lambda-p})
we find that the synchronous state remains stable for  all $p\in (0,0.5]$. Moreover, for $p>0$,
Equation~(\ref{lambda-p}) provides  a  relaxed  condition compared to the corresponding conditions
for $\lambda_0(0,m)$. In contrast, from Equation (\ref{other-q}) we find that for increasing values
of $p$ the conditions for stability become more stringent. In particular, (\ref{qq}) shows that 
a $q$-twisted state, which is stable for $p=0$, may lose stability at some positive value of $p$.
From this, we conclude that long-range random
connections promote synchrony, while inhibiting other $q$-twisted states.

Next, we discuss the stability of the synchronous regime in SW networks in a little more detail. To this 
end, we employ the leading  exponent
\be\lbl{leading}
\lambda^*(p)=\max_{m\in\N} \lambda_p(0,m)=\lambda_p(0,1), 
\ee
which reflects the rate of convergence to the synchronous state.  Larger values of $|\lambda^*(p)|$
indicate stronger synchronization. For small values of the coupling range $r>0$, from (\ref{lambda-p}),
we estimate
\be\lbl{lambda*}
\lambda^*(0)={-16\pi^2\over 3} r^3 +O(r^5).
\ee
Equation (\ref{lambda*}) shows that larger values of $r$ result in faster synchronization. Furthermore,
from (\ref{lambda-p}) and (\ref{lambda*}), we estimate the leading exponent for SW connectivity
\be\lbl{leading-SW}
\lambda^*(p)=-2p+(1-2p)\lambda^*(0).
\ee
Since for small $r>0$, $\lambda^*(0)>-1$, we have
$$
{d\over dp}\lambda^*(p)=-2(1+\lambda^*(0))<0,
$$
i.e., for increasing values of $p$ the leading exponent $| \lambda^*(p)|$ grows. 
This once again shows that adding the long-range random connections
helps synchronization in the Kuramoto model on SW graphs.

\section{Continuation}
\lbl{sec.continue}
\setcounter{equation}{0}
In this section, we develop a numerical continuation method, which will
be used to study the impact of the long-range random connections 
on the spatial structure of the stationary solutions of the Kuramoto 
model (\ref{Kuramoto}). 

\subsection{Preliminaries}
First, we adjust  our notation to the purpose of this section.
To this end, we rewrite (\ref{Kuramoto}) as follows
\be\lbl{reK}
\dot u_i =n^{-1}\sum_{j=1}^n a_{ij}(p) \sin(2\pi(u_j-u_i)),\; i\in [n],
\ee
where $u=(u_1,u_2, \dots, u_n)$ and $n\in\N$. Matrix $A(p)=(a_{ij}(p))$  is 
an $n\times n$ adjacency matrix of the SW graph $G_{n,p}$ with the randomization
parameter $p\in [0, 0.5]$. Below we will use the $\ell_1$-norm of matrix 
$A=(a_{ij})\in\R^{n\times n}$
$$
\|A\|_1=n^{-2} \sum_{i,j=1}^n |a_{ij}|.
$$

The vector form of (\ref{reK}) is given by
\be\lbl{vecK}
\dot u= F(u,p), \; 
\ee
where $F:\R^n\times \R\to \R^n$. Recall that for any $q\in\Z$, $q$-twisted state
$$
u_q = qx\pmod{1}, \; x:=(0, n^{-1}, 2n^{-1},\dots, (n-1)n^{-1})
$$ 
is a stationary solution of (\ref{vecK}) with $p=0$, i.e.,
\be\lbl{start-cont}
F(u_q,0)=0.
\ee

Thus, one can try to continue from $u_q$ for $p$ near $0$ 
to see how the addition of new random edges affects the  structure of the 
steady state solutions of (\ref{vecK}). However, one immediately 
runs into the following problem: in general, $F(\cdot, p)$
is not (stochastically) continuous in $p$, as the following lemma
shows.

\begin{lem}\lbl{lem.not-cont} Let $G_{n,p+\Delta p}$ and $G_{n,p}$ be two
independent copies of SW graphs with randomization parameters 
$0<p+\Delta p, p <0.5$. Then
\be\lbl{exp-l1}
\E \|A(p+\Delta p)-A(p)\|_1 =2\left(1-{1\over n}\right) \alpha(p),
\ee 
where 
$$
\alpha(p)=2p(1-p)+\Delta p(1-2p)=O(1),
$$
and
$A(p+\Delta p)$ and $A(p)$ stand for the adjacency matrices of 
$G_{n,p+\Delta p}$ and $G_{n,p}$ respectively.
\end{lem}
\pf
Let 
$$
L=\{(i,j)\in [n]^2:\; 0<\min\{|i-j|, n-|i-j|\} \le k\} 
$$ 
be the set of indices corresponding to the edges of $k$-NN graph, and
let $L^c=[n]^2/L$ be its complement. 

Denote RVs
$$
\xi_{ij}=|a_{ij}(p+\Delta p)-a_{ij}(p)|,\; 1\le i<j\le n.
$$
We show that $\{\xi_{ij}\} $ are IID binomial RVs. 
Indeed, let $1\le i<j\le n$ and $(i,j)\in L$. Then
\begin{eqnarray*}
\P\{\xi_{ij}=1\}&=& \P\left\{ \left(a_{ij}(p+\Delta p)=1\right)\& \left(a_{ij}(p)=0\right)\right\}
+ \P\left\{ \left(a_{ij}(p+\Delta p)=0\right)\& \left(a_{ij}(p)=1\right)\right\}\\
&=& (1-p-\Delta p)p+(p+\Delta p) (1-p)=\alpha(p).
\end{eqnarray*}

Likewise, for $1\le i<j\le n$ and $(i,j)\in L^c,$ we have 
\begin{eqnarray*}
\P\{\xi_{ij}=1\}&=& \P\left\{ \left(a_{ij}(p+\Delta p)=1\right)\& \left(a_{ij}(p)=0\right)\right\}
+ \P\left\{ \left(a_{ij}(p+\Delta p)=0\right)\& \left(a_{ij}(p)=1\right)\right\}\\
&=& (p+\Delta p) (1-p)+(1-p-\Delta p)p=\alpha(p).
\end{eqnarray*}

Thus, $\mathcal{L}(\xi_{ij})=\operatorname{Bin}(\alpha(p)), 1\le i<j\le n,$ and
$$
\E\|A(p+\Delta p)-A(p)\|_1={2\over n^2}
\E\left(\sum_{1\le i<j\le n} \xi_{ij} \right)={2n(n-1)\over n^2} \alpha(p).
$$
$\qed$

\begin{rem} Note that for $p\in (0,1/2)$,  $\E\|A(p+\Delta p)-A(p)\|_1$ stays bounded away from $0$ as
$\Delta p\to 0$.
\end{rem}
\begin{rem} Using the strong law of large numbers, one can show that
\be\lbl{almost-surely}
\lim_{n\to\infty} \left\|A(p+\Delta p)-A(p)\right\|_1 =2\alpha(p)\quad\mbox{almost surely}.
\ee 

Indeed, since $\{\xi_{ij}\}_{1\le i<j \le n}$ are IID binomial RVs, $\mathcal{L}(\xi_{12})=\operatorname{Bin}(\alpha(p))$,
by the strong law of large numbers, we have
\be\lbl{apply-SLLN}
\P\left\{ \lim_{n\to\infty} {1\over n(n-1)} \sum_{1\le i<j\le n} \xi_{ij} =\alpha (p)\right\} =1.
\ee
Then (\ref{almost-surely}) follows from (\ref{apply-SLLN}), because
$$
\left\|A(p+\Delta p)-A(p)\right\|_1=2\left(1-{1\over n}\right) {1\over n(n-1)} \sum_{1\le i<j\le n} \xi_{ij}.
$$
\end{rem}
\begin{figure}
{\bf a}\;\;\includegraphics[height=1.6in,width=1.8in]{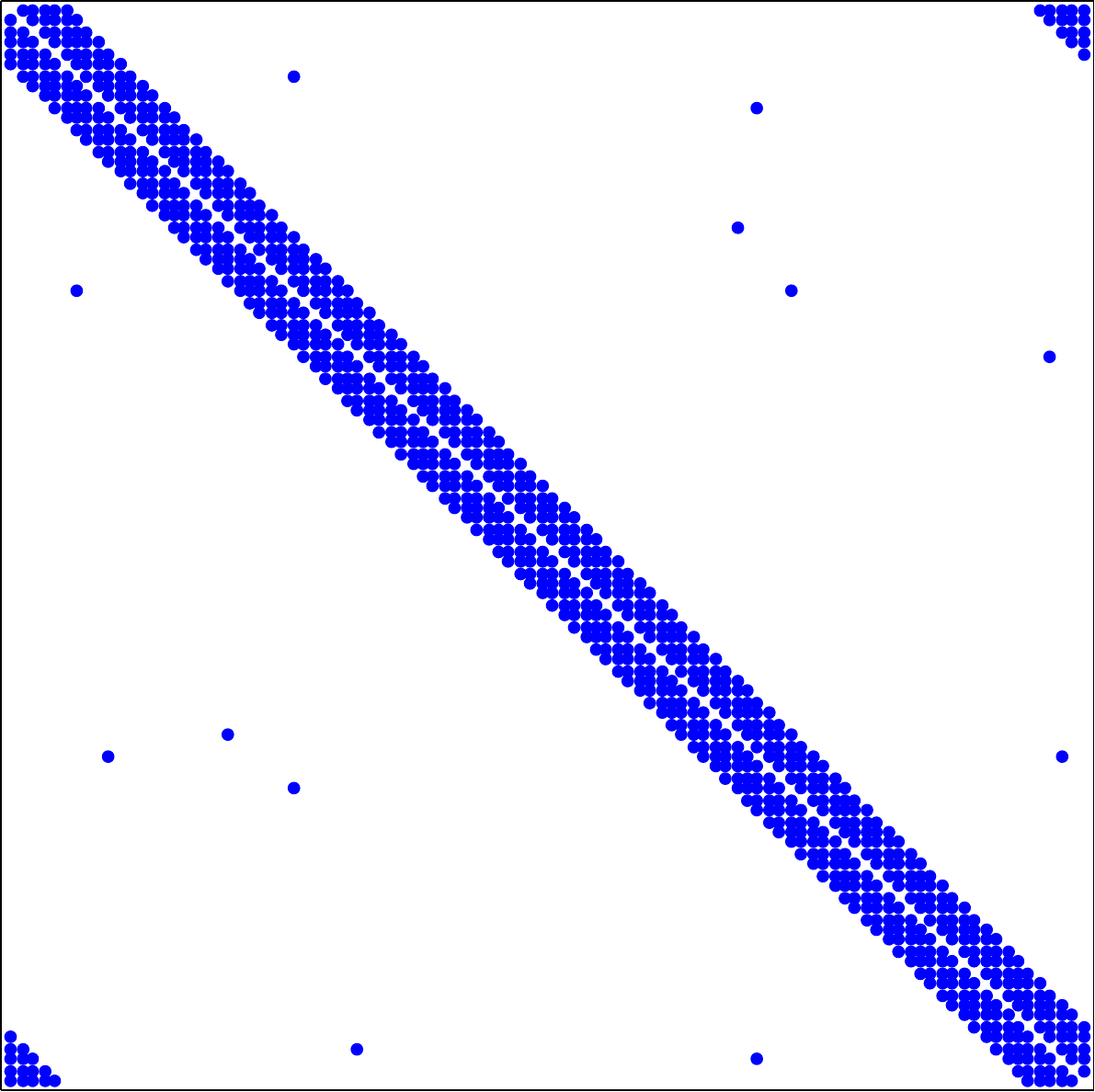}\quad
{\bf b}\;\;\includegraphics[height=1.6in,width=1.8in]{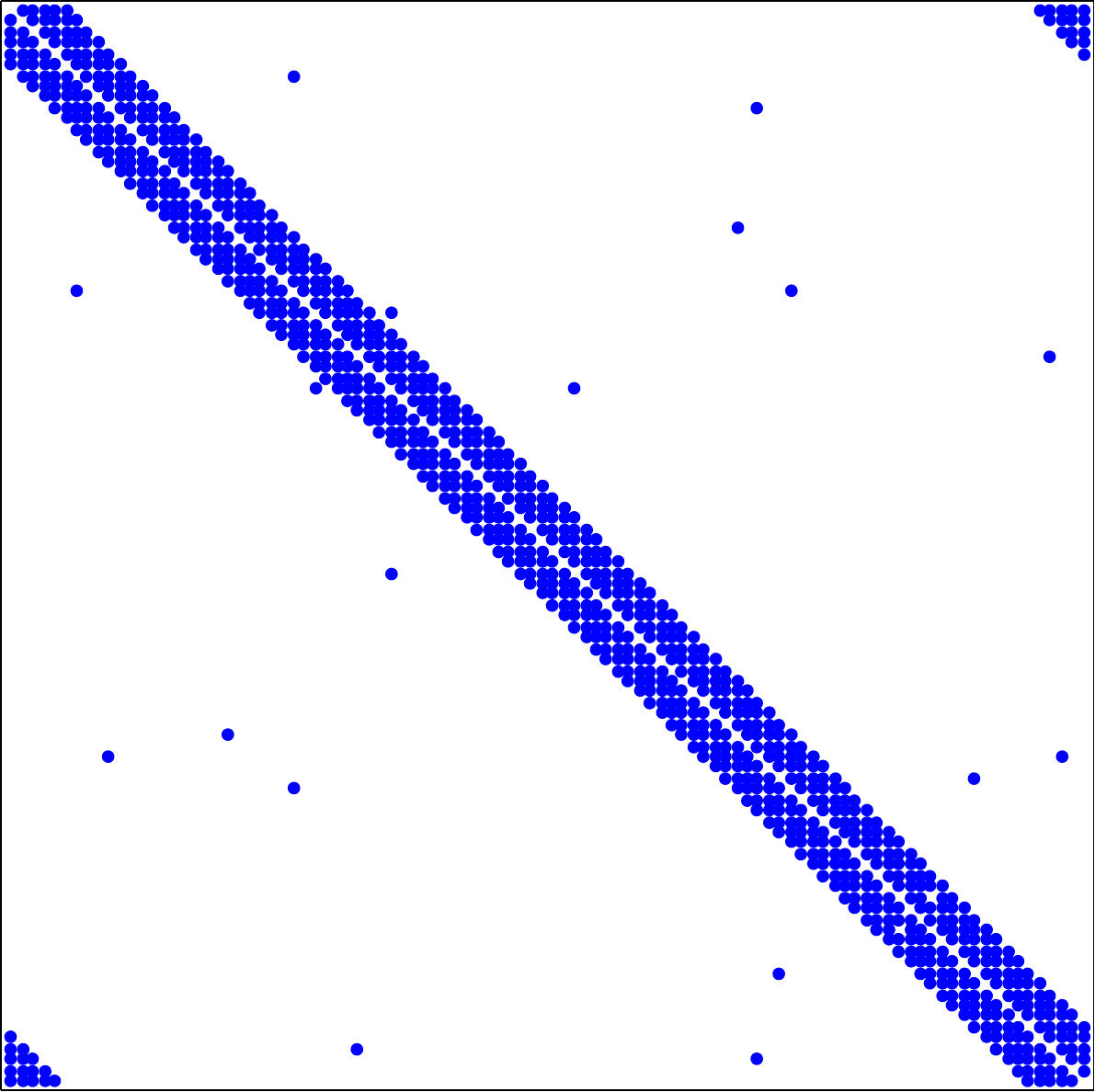}\quad
{\bf c}\;\;\includegraphics[height=1.6in,width=1.8in]{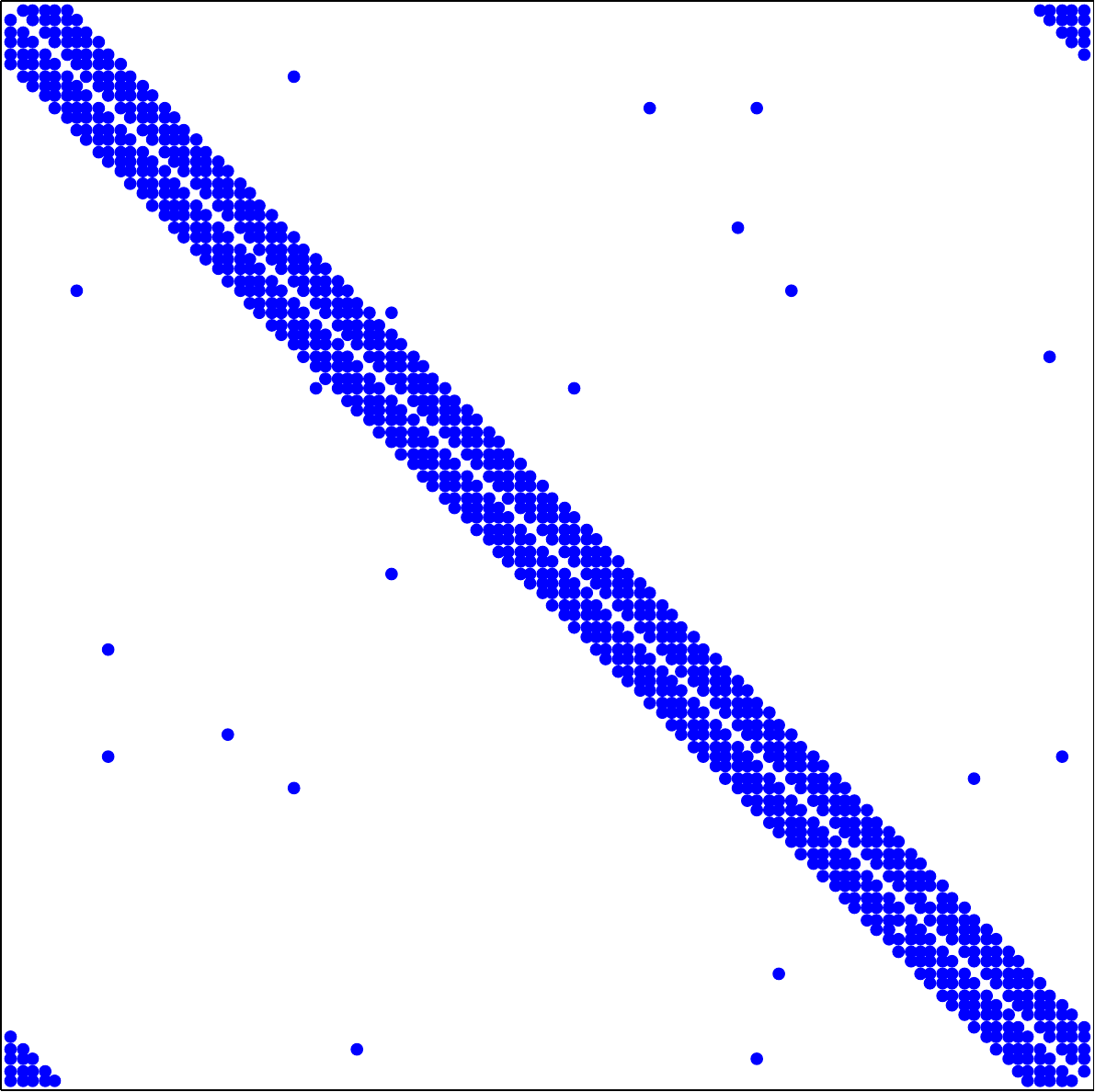}
\caption{Schematic representation of the adjacency matrices of three consistent SW graphs 
$G_{n,p}$ generated using (\ref{dp1})-(\ref{dp4}): $n=100$, $p=0.001$ (\textbf{a}), 
$p=0.002$ (\textbf{b}),  $p=0.003$ (\textbf{c}). Black dots denote  entries of the adjacency 
matrix that are equal to $1$.
} 
\lbl{f.1}
\end{figure}  
\subsection{Consistent family of SW graphs}

Lemma~\ref{lem.not-cont} shows that $F(u,p)$ is not stochastically continuous in $p$.
Therefore, nonlinear equation
\be\lbl{nonlinear}
F(u,p)=0,
\ee
as it stands can not be used for continuation. To rectify this problem, we introduce a
special family of SW graphs, along which $F(u,p)$ is stochastically  continuous in $p$.

\begin{df}
Let $N\gg 1$ and $h=2N^{-1}$. Consider a family of SW graphs $\{G_{n,p_k}\}_{k=0}^N$, 
where $p_k=kh$. Graphs $G_{n,p_k}$ are said to form a consistent family of SW graphs if 
\be\lbl{consitent}
 \E \|A(p_{k+1})-A(p_k)\|_1 =O(h)\qquad \forall k\in\{0,1,2,\dots,N-1\}.
\ee
\end{df}

It is not hard to construct a consistent family of SW graphs. This can be done inductively.
Suppose $G_{n,p_k}$ is given for some $0\le k\le N-1$. Then $G_{n,p_{k+1}}$ is
a random graph defined by the following conditions.
For $1\le i<j\le n$ and $(i,j)\in L$, we set
\begin{eqnarray}\lbl{dp1}
\P\{ a_{ij}(p_{k+1})=0 | a_{ij}(p_k)=1\} &=& q_k, \; q_k:=h(1-p_k)^{-1},\\
\lbl{dp2}
\P\{ a_{ij}(p_{k+1})=0 | a_{ij}(p_k)=0\} &=& 1.
\end{eqnarray}
Likewise, for $1\le i<j\le n$ and $(i,j)\in L^c$, we have
\begin{eqnarray}\lbl{dp3}
\P\{ a_{ij}(p_{k+1})=1 | a_{ij}(p_k)=1\}&=&1,\\
\lbl{dp4}
\P\{ a_{ij}(p_{k+1})=1 | a_{ij}(p_k)=0\}&=&q_k.
\end{eqnarray}
It is easy to check that $A(p_{k+1})$ defines a SW graph $G_{n,p_{k+1}}$.
Moreover, graphs $G_{n,p_k}$ constructed above form a consistent
family.

\begin{lem}\lbl{lem.consistent} Recursive relations  (\ref{dp1})-(\ref{dp4}) define
 a consistent family of SW graphs.
\end{lem}
\pf\;
Let $k\in \{0,1,\dots,N-1\}$ be arbitrary and define
$$
\xi_{ij}=|a_{ij}(p_{k+1})-a_{ij}(p_k)|,\; 1\le i<j\le n.
$$
We show that $\mathcal{L}(\xi_{ij})=\operatorname{Bin}(h)$. 
Indeed, let $(i,j)\in L$. Then
\begin{eqnarray*}
\P\{\xi_{ij}=1\}&=& \P\{ a_{ij}(p_{k+1})\neq a_{ij}(p_k) \}=
\P\{ (a_{ij}(p_{k+1})=0) \& (a_{ij}(p_k)=1) \}\\
                  &=& \P\{ a_{ij}(p_{k+1})=0| a_{ij}(p_k)=1\} \P\{a_{ij}(p_k)=1 \}= q_k(1-p_k)=h.
\end{eqnarray*}
Likewise, for $1\le i<j\le n$ and $(i,j)\in L^c$, we have
\begin{eqnarray*}
\P\{\xi_{ij}=1\}&=& \P\{ a_{ij}(p_{k+1})\neq a_{ij}(p_k) \}=
\P\{ (a_{ij}(p_{k+1})=1) \& (a_{ij}(p_k)=0) \}\\
                  &=& \P\{ a_{ij}(p_{k+1})=1| a_{ij}(p_k)=0\} \P\{a_{ij}(p_{k+1})=0 \}= q_k(1-p_k)=h.
\end{eqnarray*}
Similarly, we compute $\P\{\xi_{ij}=1\}=1-h$, $1\le i< j \le n$. Thus,
$$
\E \|A(p_{k+1})-A(p_k)\|_1= {2n(n-1)\over n^2} h=2(1-n^{-1})h.
$$
$\qed$
\begin{figure}
{\bf a}\includegraphics[height=1.4in,width=1.5in]{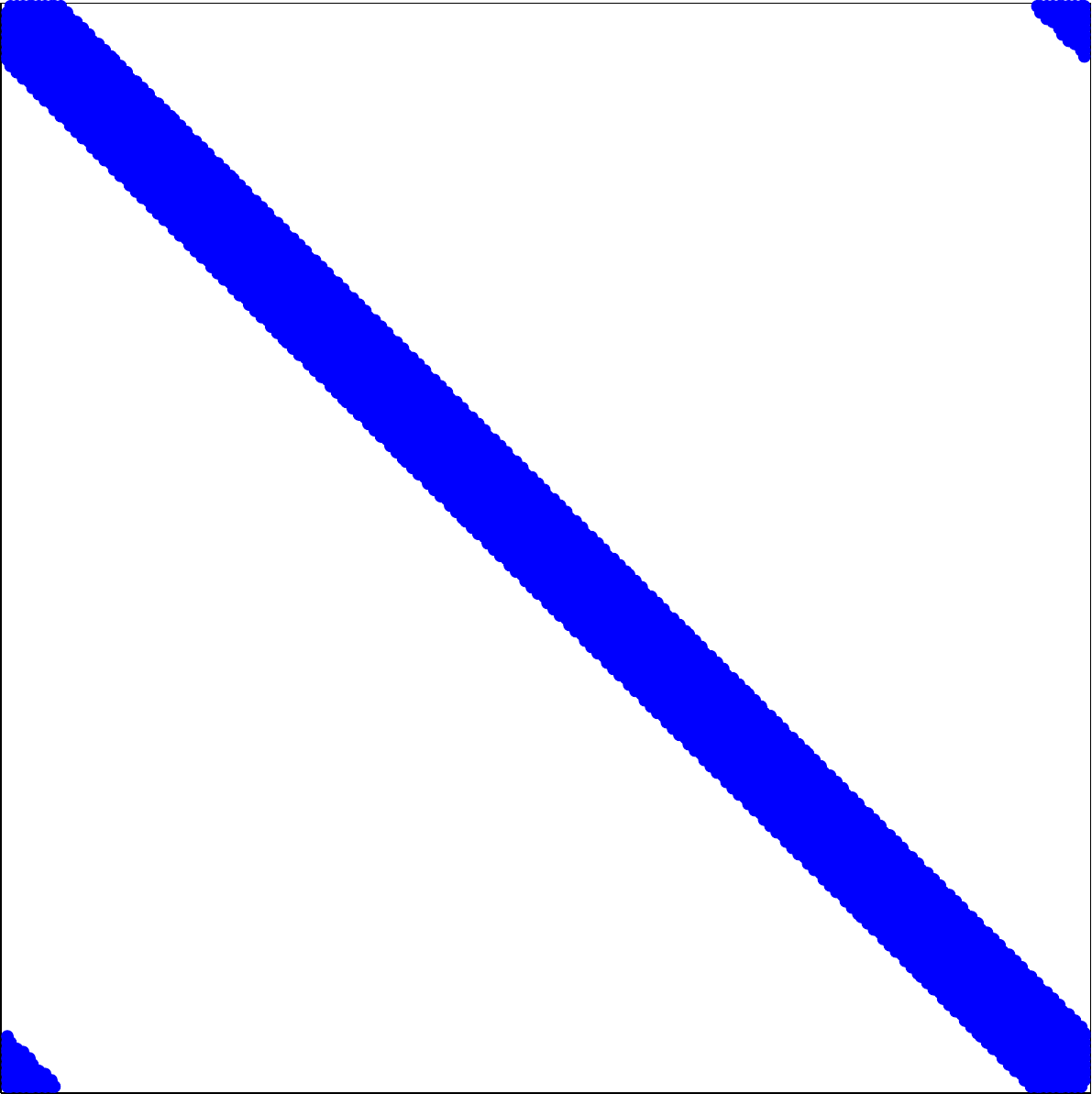}
{\bf b}\includegraphics[height=1.4in,width=1.5in]{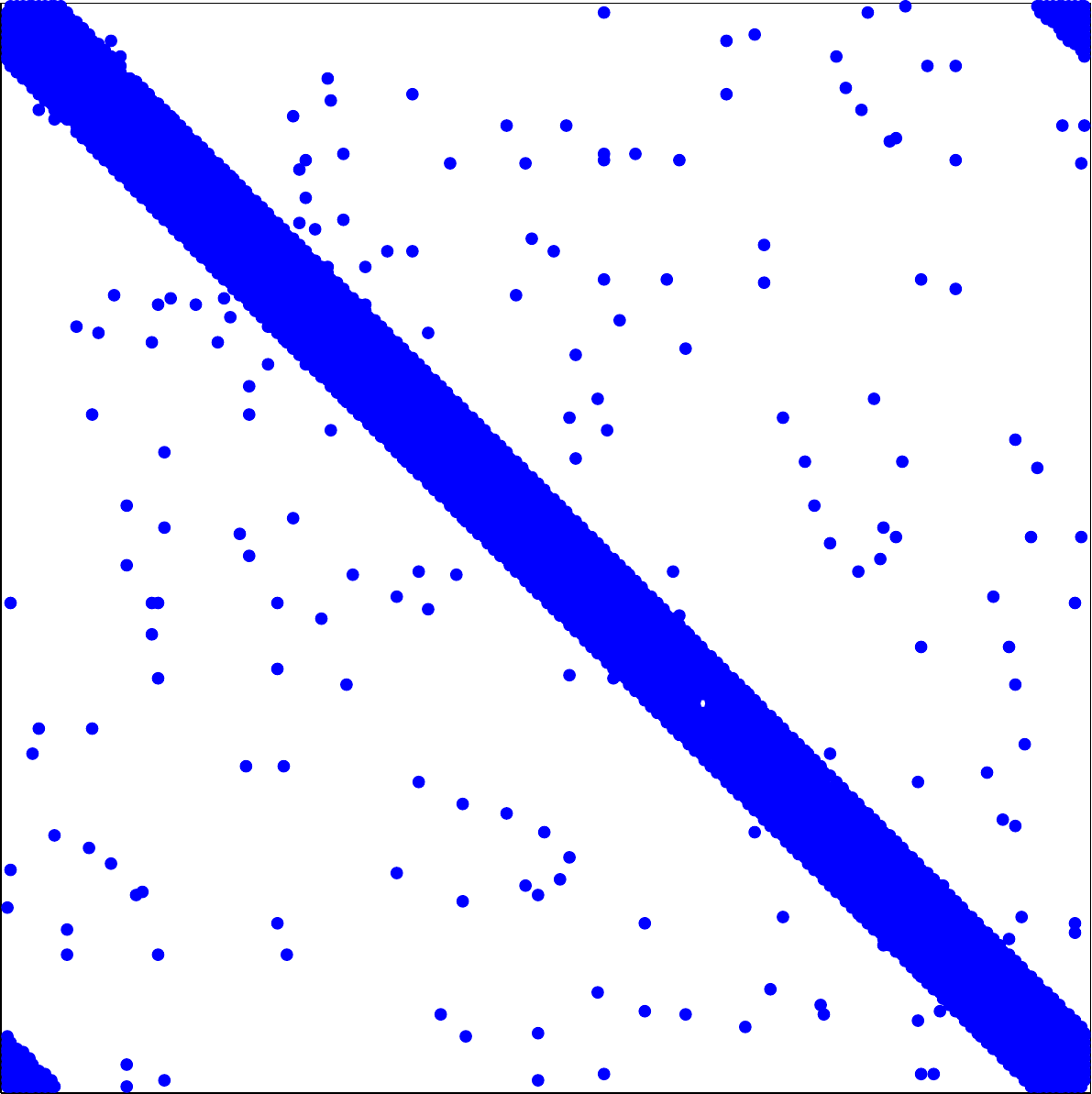}
{\bf c}\includegraphics[height=1.4in,width=1.5in]{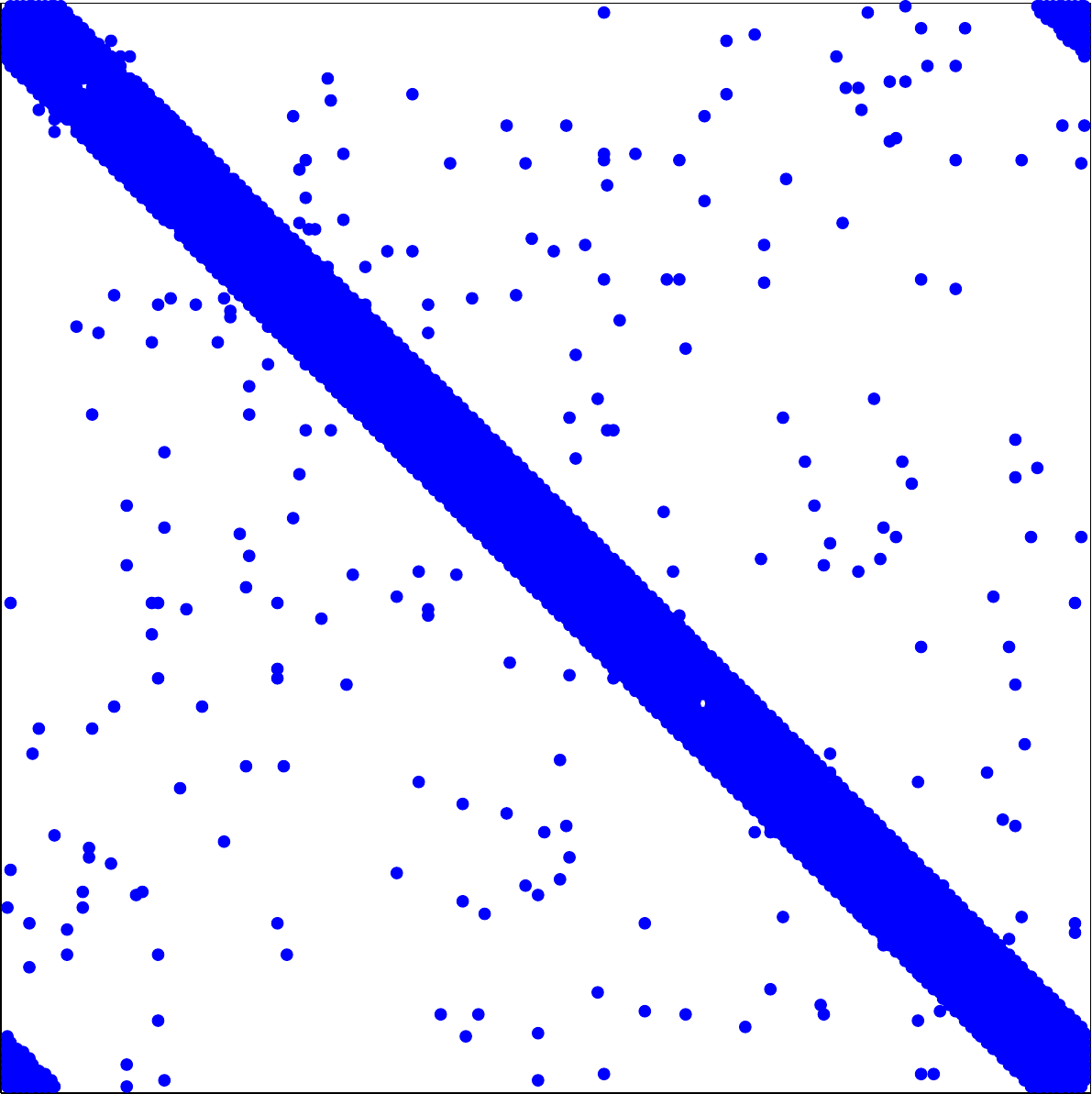}
{\bf d}\includegraphics[height=1.4in,width=1.5in]{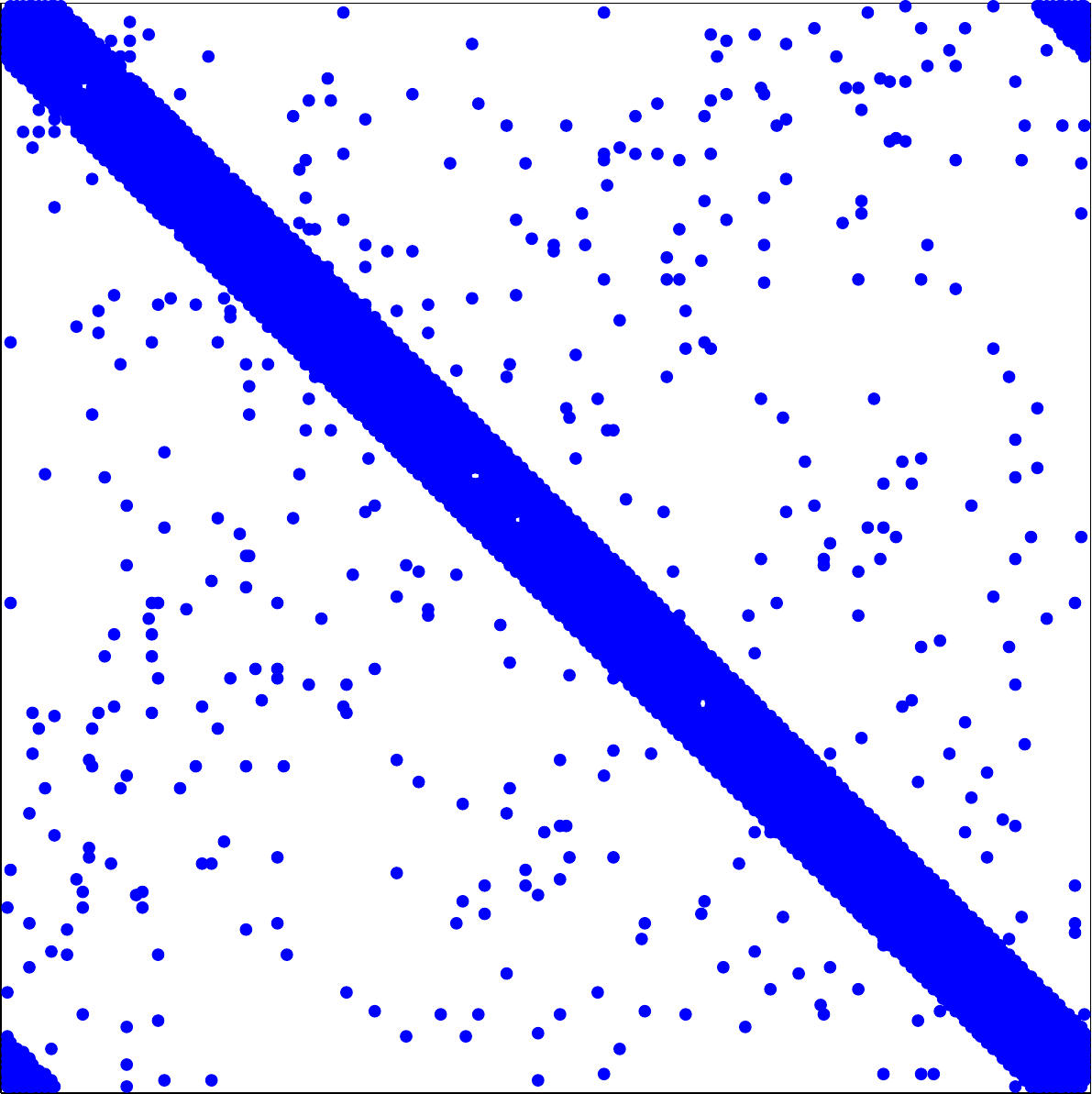}\\
{\bf e}\includegraphics[height=1.4in,width=1.5in]{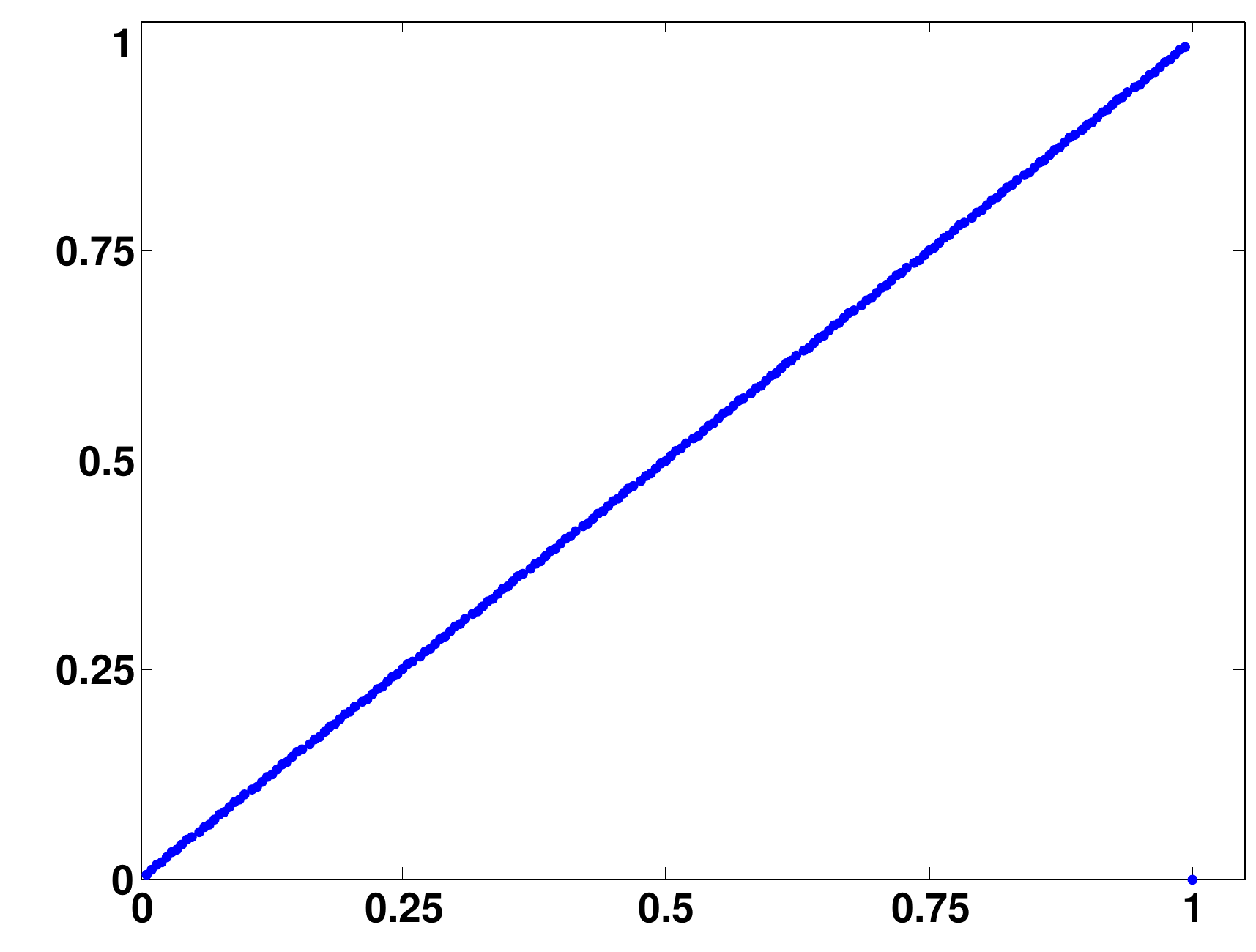}
{\bf f}\includegraphics[height=1.4in,width=1.5in]{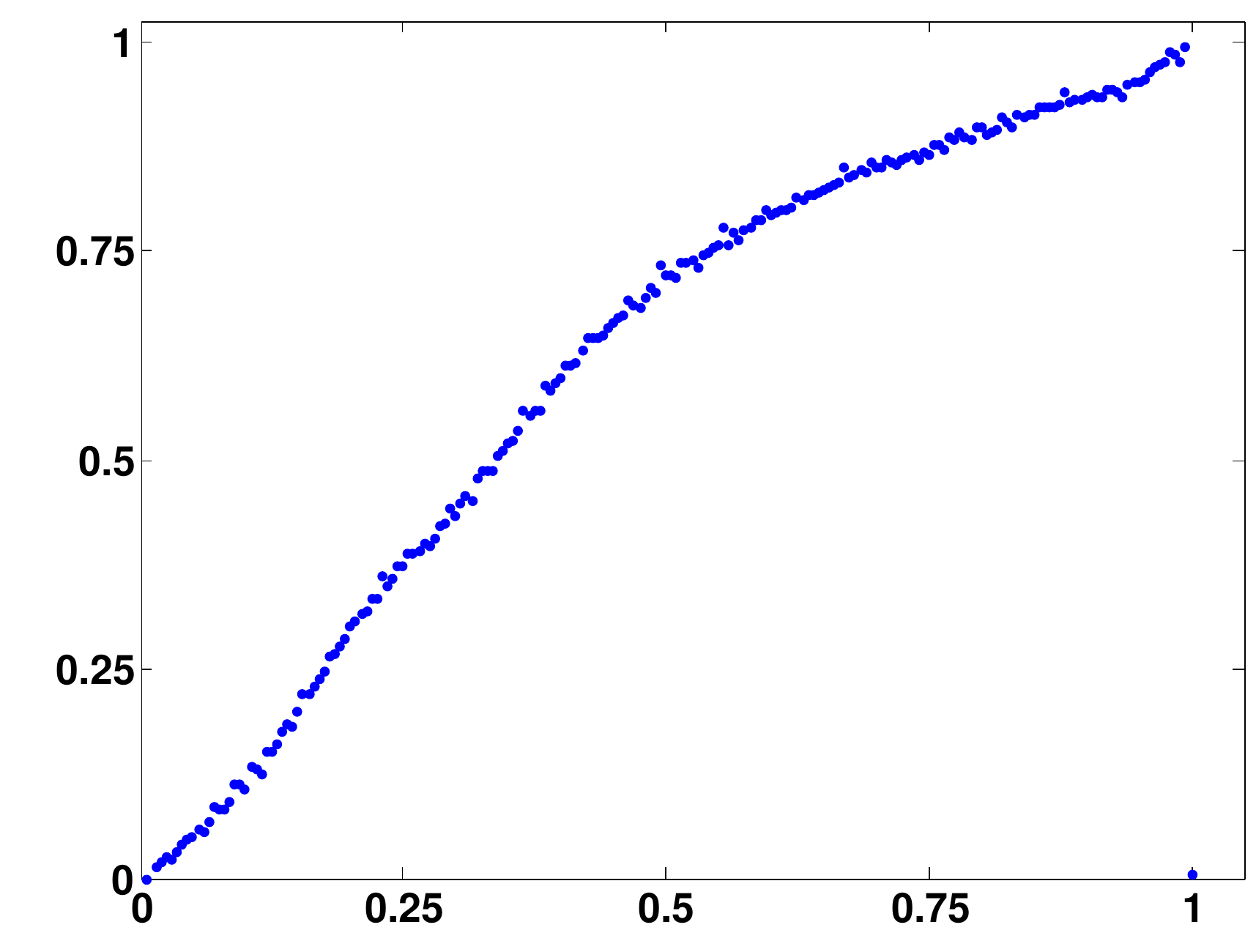}
{\bf g}\includegraphics[height=1.4in,width=1.5in]{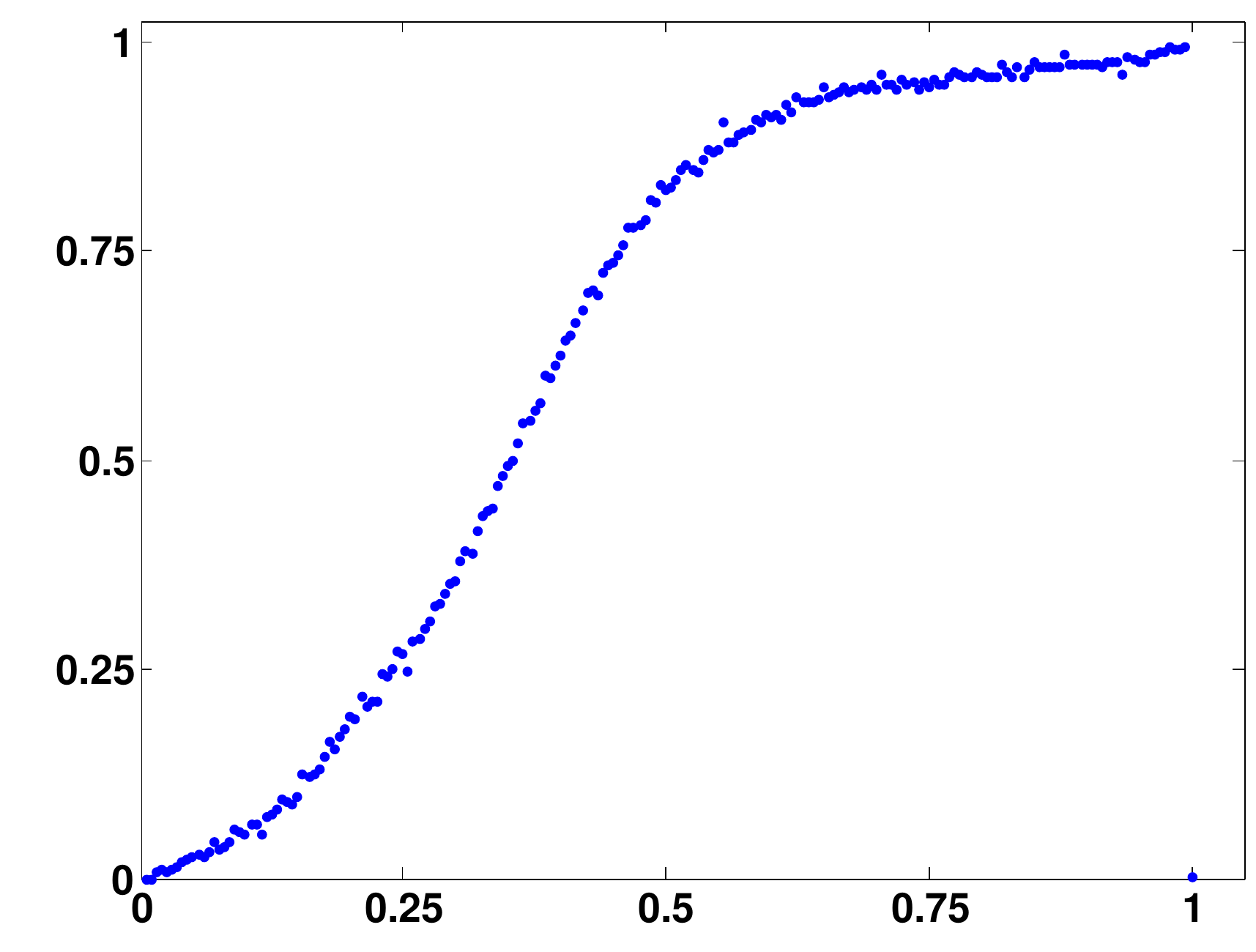}
{\bf h}\includegraphics[height=1.4in,width=1.5in]{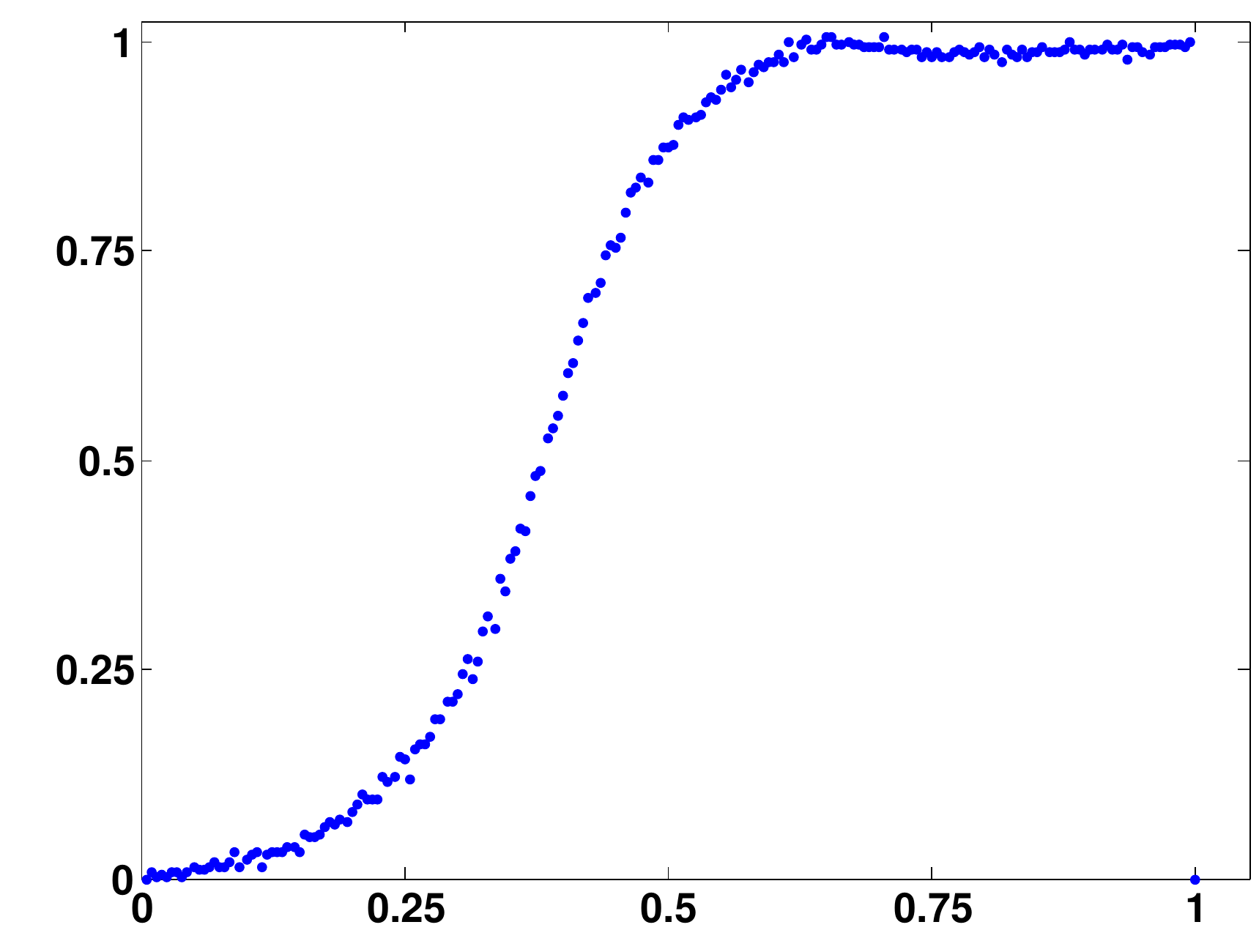}
\caption{A consistent family of the SW graphs $G_{100,p}$ ({\bf a}-{\bf d}).
The corresponding steady-state solutions (\ref{reK}) obtained by continuation
from $1$-twisted state solution ({\bf e}) are shown in ({\bf f}-{\bf h}).
The corresponding values of $p$ are $3.5\cdot 10^{-3}$ ({\bf f}),
$4.9\cdot 10^{-3}$ ({\bf g}), and $9.9\cdot 10^{-3}$ ({\bf h}).
}
\lbl{f.2}
\end{figure}  

Figure~\ref{f.1} presents several representative examples of SW graphs 
from a consistent family, which where computed using
(\ref{dp1})-(\ref{dp4}). Note that the adjacency matrices in Fig.~\ref{f.1}
change little for small changes in the randomization parameter $p$.
\begin{figure}
{\bf a}\includegraphics[height=1.4in,width=1.5in]{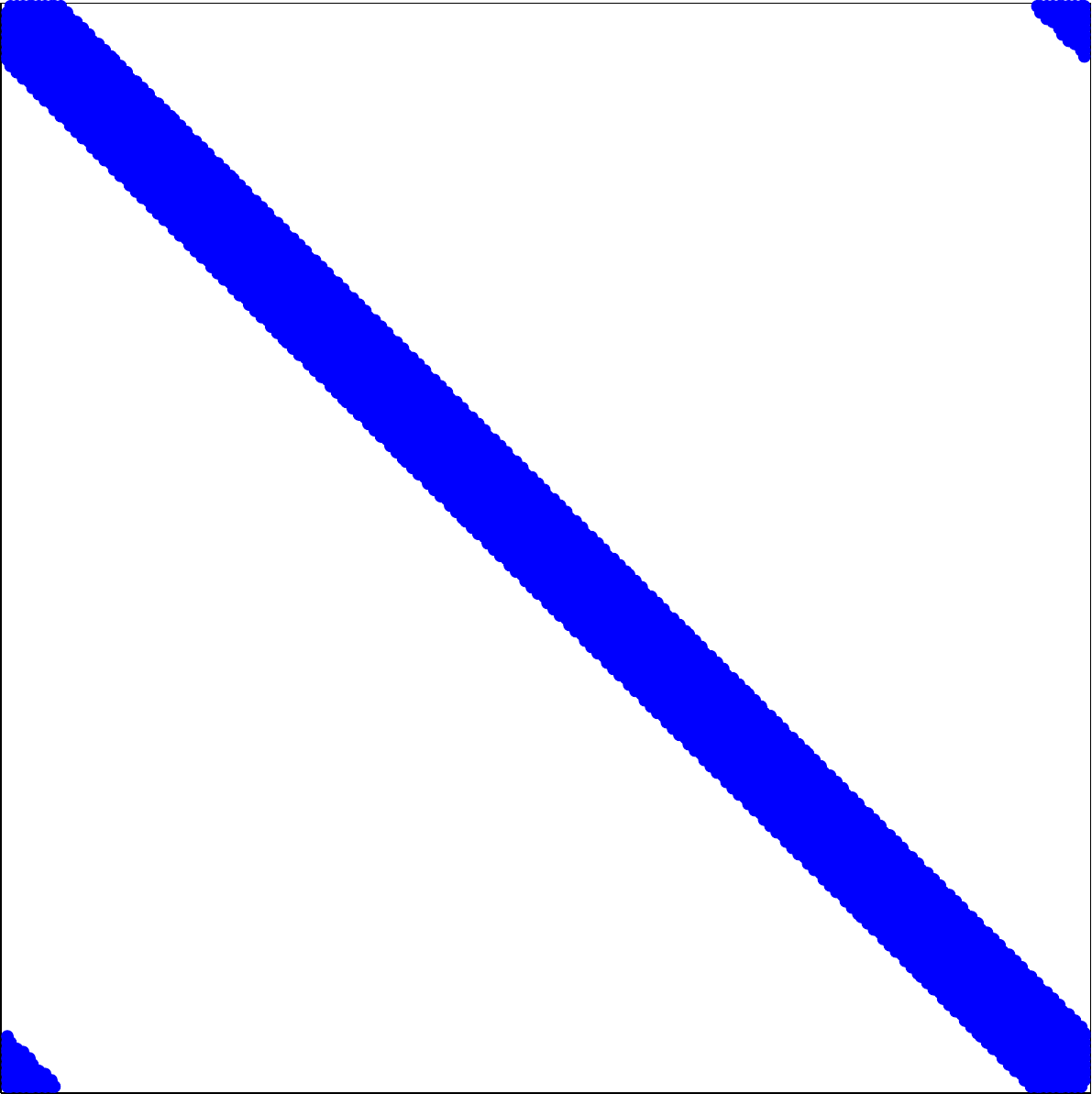}
{\bf b}\includegraphics[height=1.4in,width=1.5in]{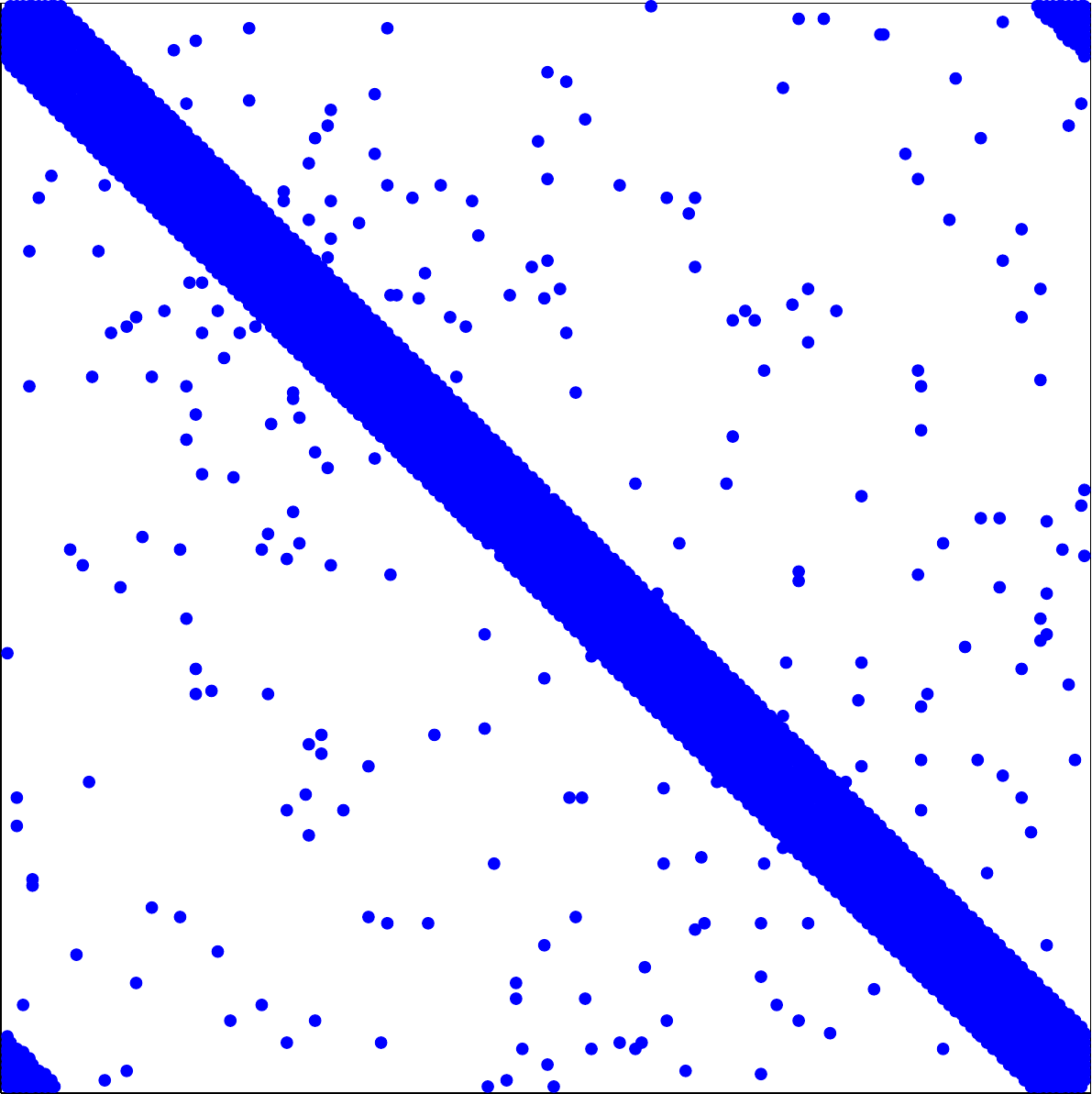}
{\bf c}\includegraphics[height=1.4in,width=1.5in]{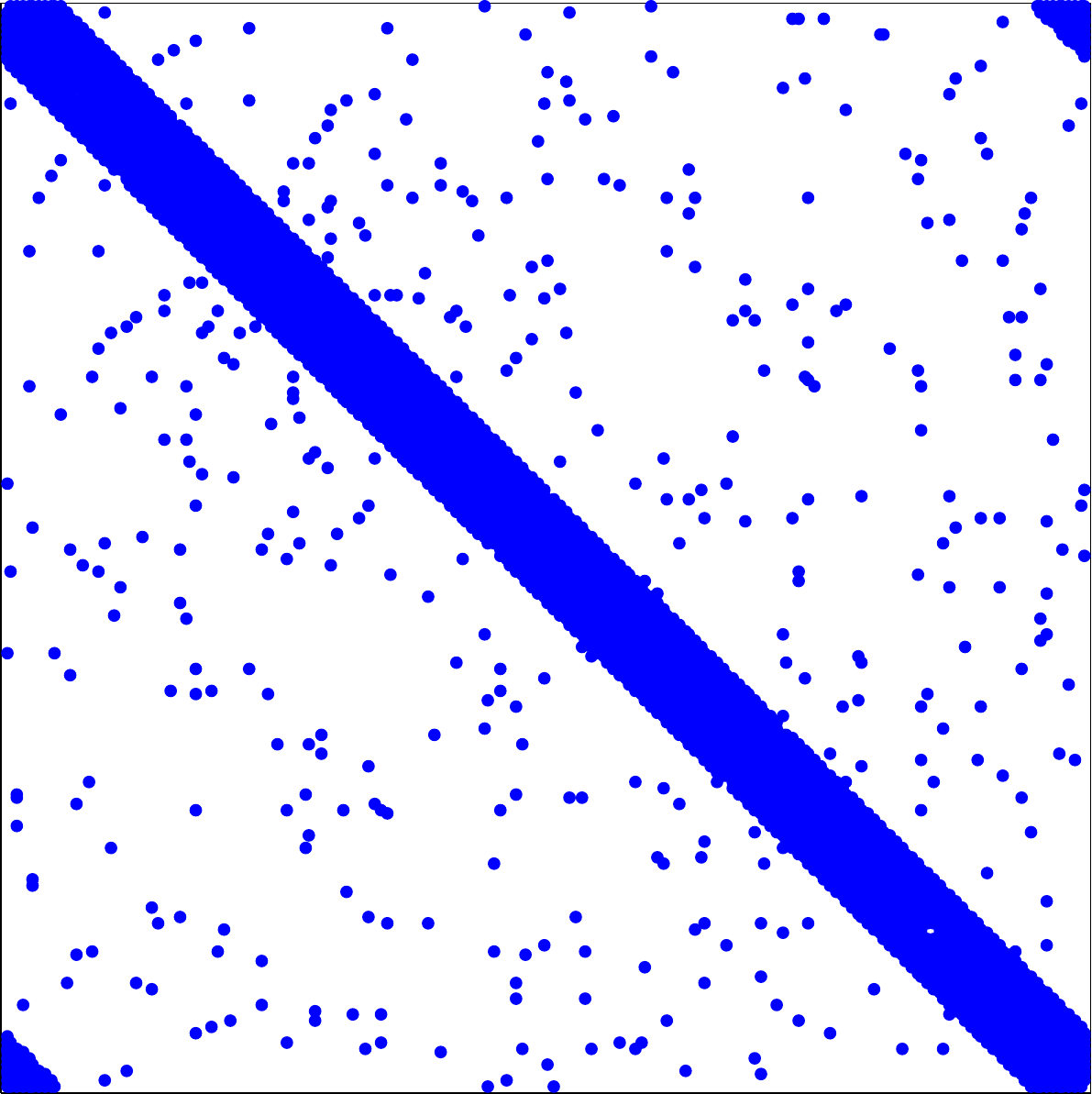}
{\bf d}\includegraphics[height=1.4in,width=1.5in]{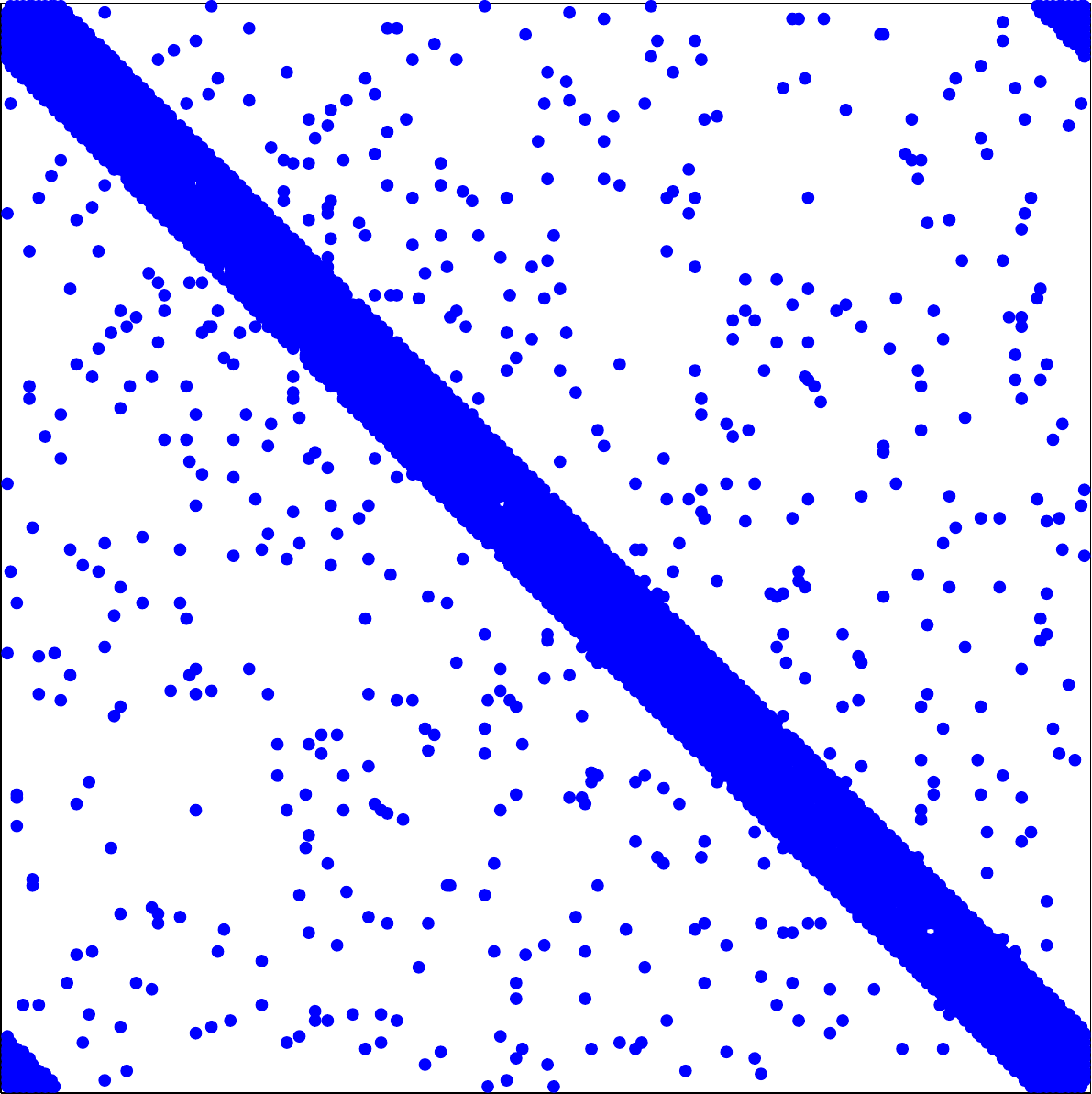}\\
{\bf e}\includegraphics[height=1.4in,width=1.5in]{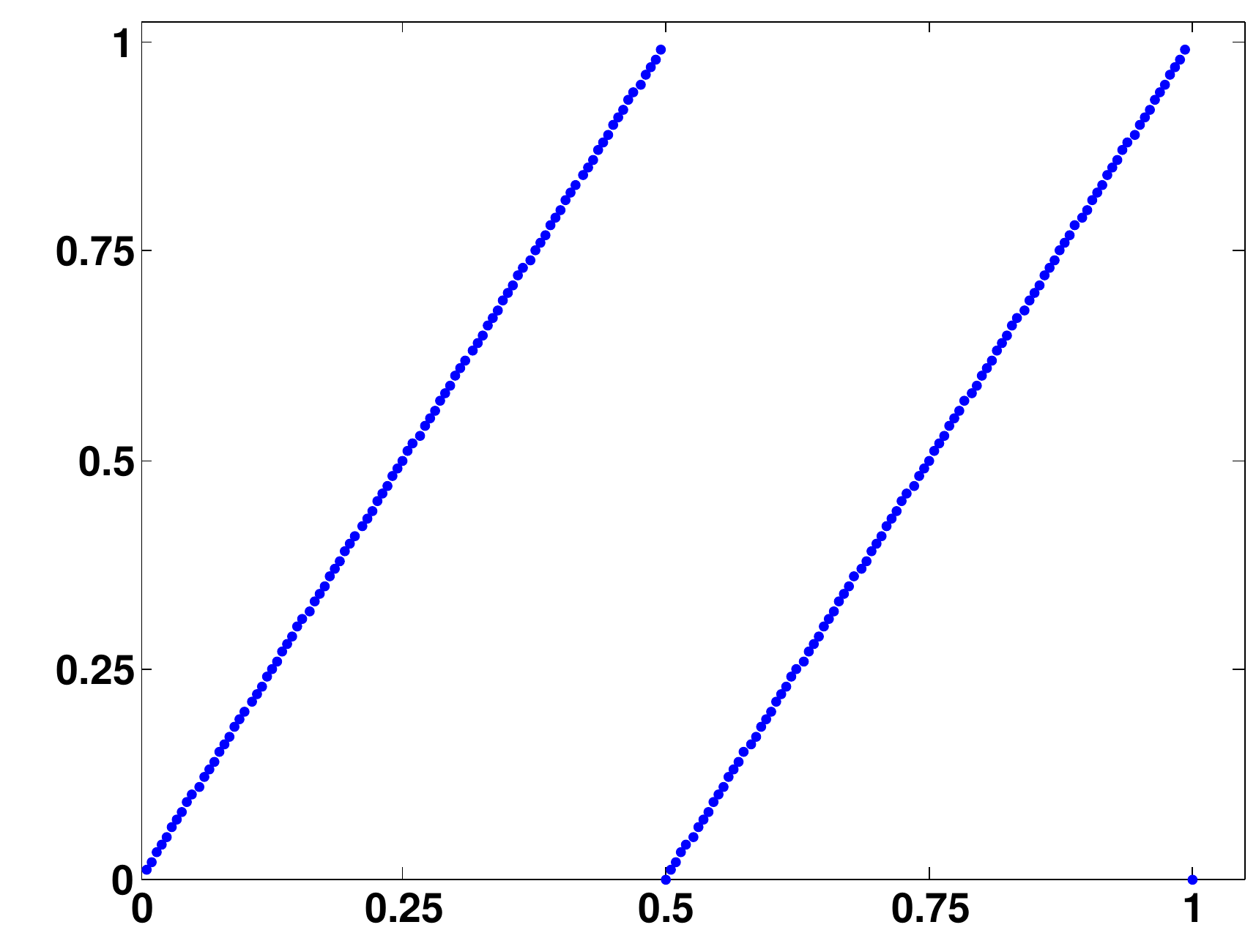}
{\bf f}\includegraphics[height=1.4in,width=1.5in]{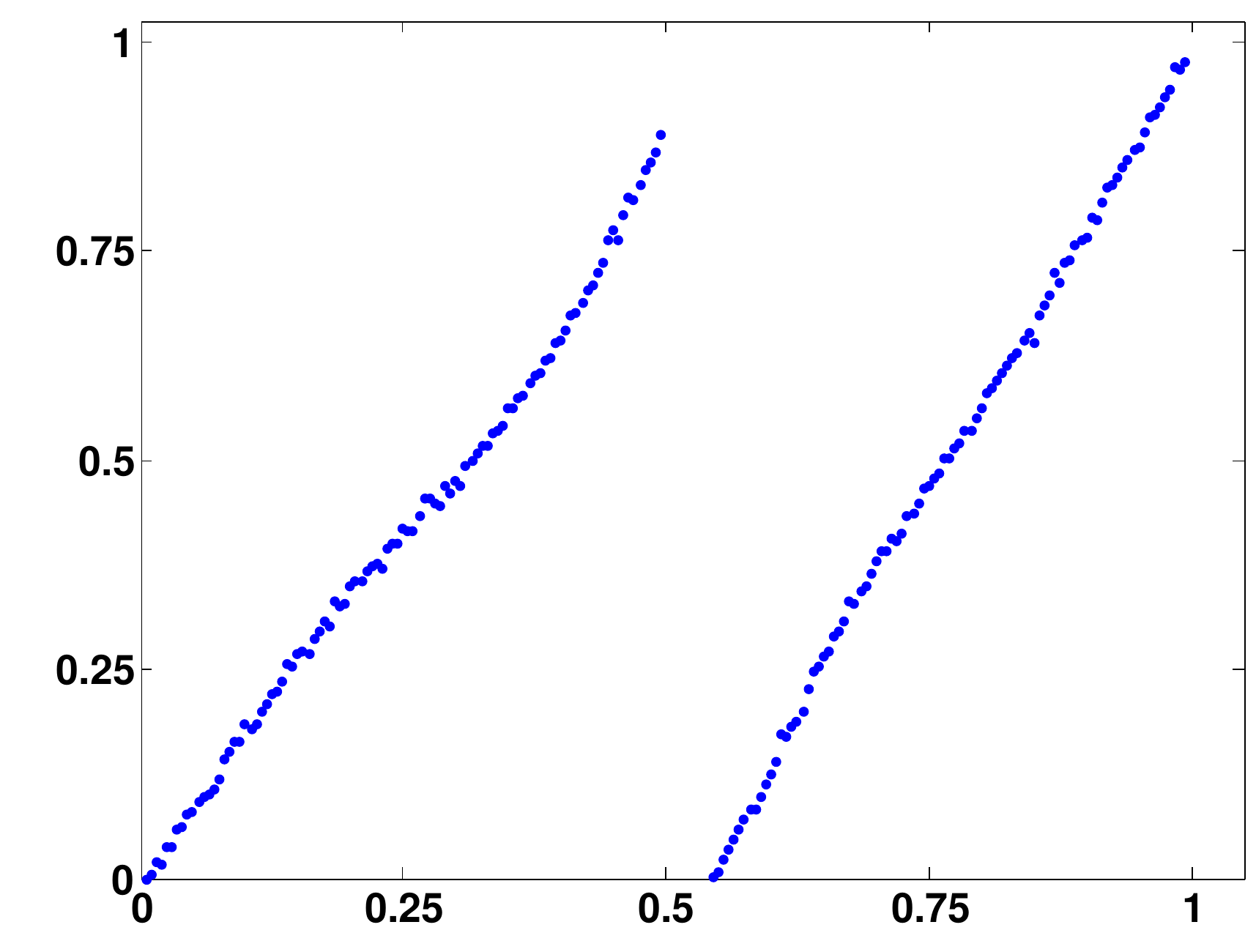}
{\bf g}\includegraphics[height=1.4in,width=1.5in]{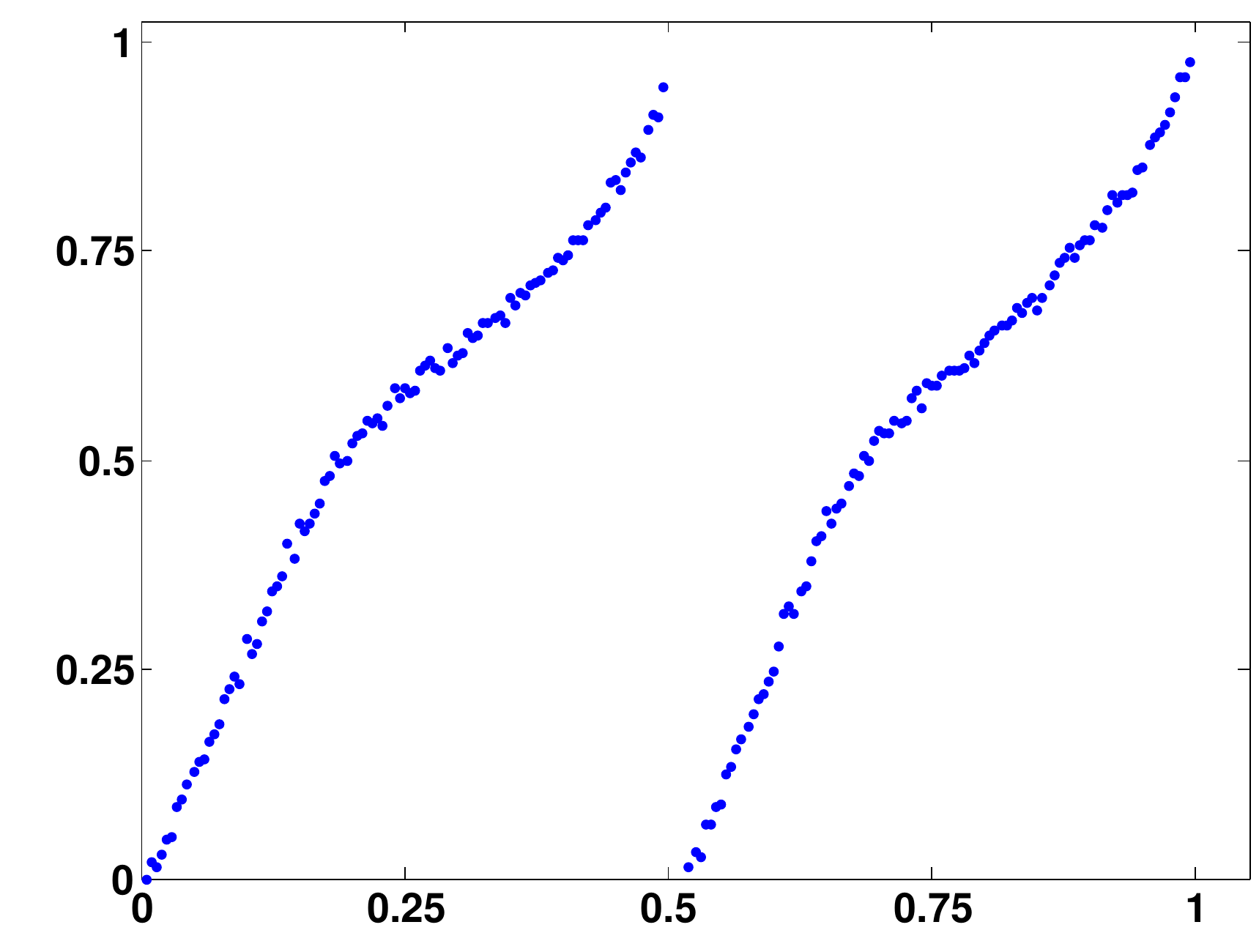}
{\bf h}\includegraphics[height=1.4in,width=1.5in]{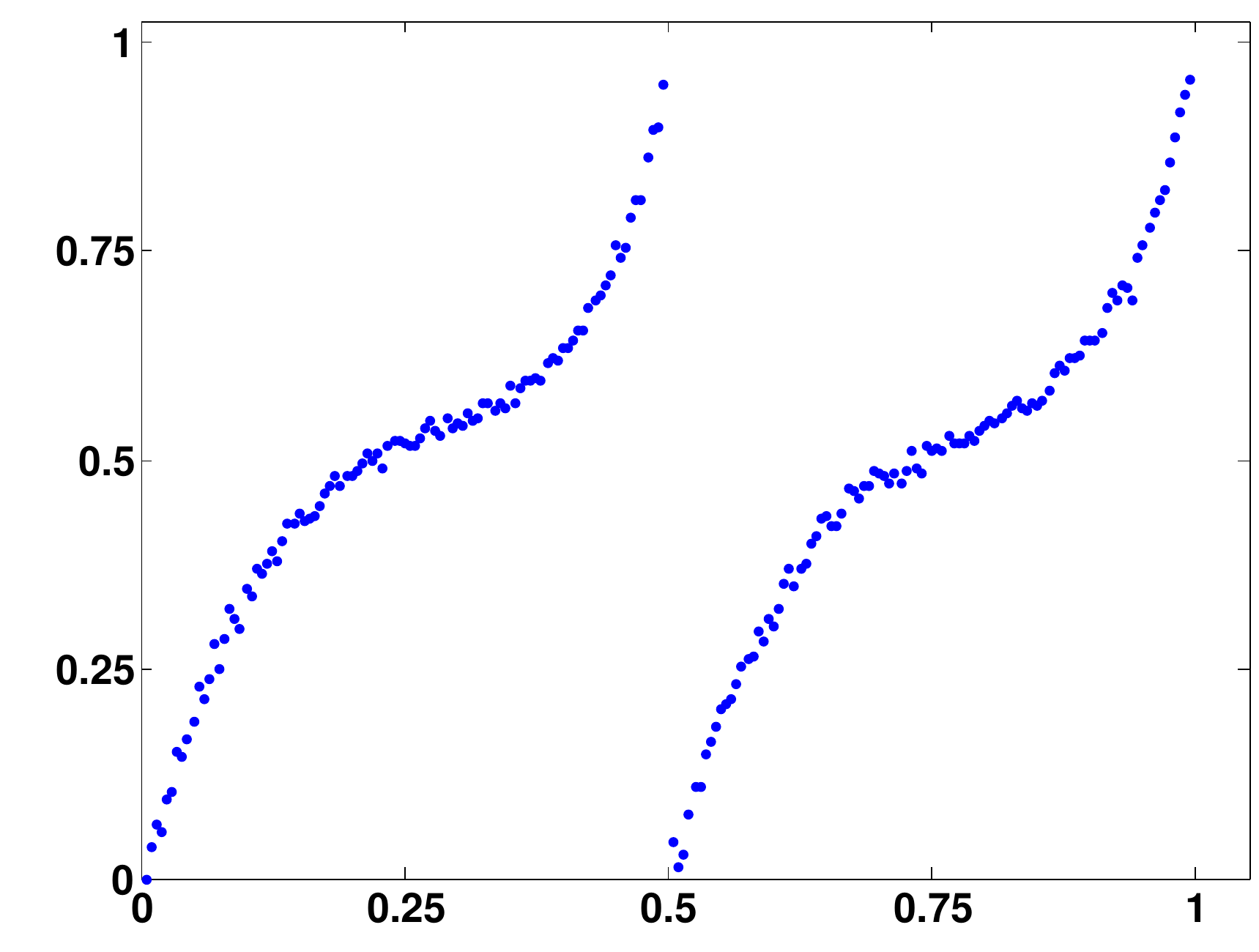}
\caption{A consistent family of the SW graphs $G_{200,p}$ ({\bf a}-{\bf d}).
The corresponding steady-state solutions (\ref{reK}) obtained by continuation
from $2$-twisted state solution ({\bf e}) are shown in ({\bf f}-{\bf h}).
The corresponding values of $p$ are $6\cdot 10^{-3}$ ({\bf f}),
$1.1\cdot 10^{-3}$ ({\bf g}), and $1.63\cdot 10^{-2}$ ({\bf h}).
}
\lbl{f.3}
\end{figure}

\subsection{The Newton's continuation and numerical results}

With the idea of the consistent family of SW graphs at hand, we can now
set up a numerical continuation procedure based on Newton's method.

Let $h>0$ be sufficiently small and $\{G_{n,p_k}\}_{k=0}^N, \, N=\lfloor (2h)^{-1}\rfloor$ be a consistent family 
of SW graphs (cf. (\ref{dp1})-(\ref{dp4})). Then the right hand side of 
(\ref{reK}) depends continuously on $p$. Indeed, from (\ref{reK})
using the definition of the $\ell_1$-norm we have
\be\lbl{easy-cont}
\|F(u,p_{k+1})-F(u,p_k)\|_1\le \|A(p_{k+1})-A(p_k)\|_1,
\ee
where  by $\ell_1$-norm of vector $F=(F_1,F_2,\dots,F_n)$ we mean
$$
\|F\|_1=n^{-1}\sum_{i=1}^n |F_i|.
$$
From (\ref{easy-cont}) and Lemma~\ref{lem.consistent}, we have
$$
\E\|F(u,p_{k+1})-F(u,p_k)\|_1\le \E\|A(p_{k+1})-A(p_k)\|_1=O(h)
$$
uniformly in $u\in\R^n$.

To compute $\tilde u(p_{k+1})$, the approximation of  $u(p_{k+1})$, the steady state
solution of the Kuramoto model (\ref{reK}) on the SW graph $G_{n,p_{k+1}}$,
 we use Newton's iterations
\be\lbl{Newton}
u^{(k+1)}=u^{(k)}-\left[ {\p\over\p u} F(u(p),p+\Delta p)\right]^+  F(u^{(k)},p+\Delta p),
\ee
until the error $\|u^{(k+1)}-u^{(k)}\|$ gets sufficiently small. To initialize
(\ref{Newton}), we use  $u^{(0)}=\tilde u(p_k)$, the approximation
of the steady-state solution of (\ref{reK}) on $G_{n,p_k}$, which was computed at the previous step
of the continuation algorithm. In (\ref{Newton}),  the Moore-Penrose pseudo-inverse of the Jacobian matrix 
$\left[ {\p\over\p u} F(u(p),p)\right]^+$ is used, because  
$$
\operatorname{ker} {\p\over\p u} F(u(p),p)=\operatorname{span}\{ (1,1,\dots,1)^\t\}.
$$
The singularity of the Jacobian matrix is due to the translational
invariance of solutions of (\ref{reK}).

The results of the numerical continuation from $q$-twisted states are
shown in Figs~\ref{f.2} and \ref{f.3}.
They highlight the importance of the structure of the network in shaping
the spatial patterns. Note that the change of approximately $1\%$ of the 
entries in the $\{0,1\}$-valued adjacency matrix was sufficient  to transform 
the  $1$-twisted state shown in Fig.~\ref{f.2}e to the
step-like solution in Fig.~\ref{f.2}h. 
The numerics also reveal an unexpected effect on the spatial organization of the steady states of the Kuramoto model
caused by replacing local connections with long-range random ones.
It is best seen from the numerical experiments for $1$-twisted states shown in Fig.~\ref{f.2}.
Note that for increasing values of $p$, the spatial profile develops 
an  interface with  progressively increasing slope. We observed the same result for all 
random realizations of the consistent families of SW networks that we used in these experiments.
Furthermore, this effect persists for $q$-twisted states with larger values of $q$. For example, the solution continued from the
$2$-twisted state shown in Fig.~\ref{f.3} has two plateaus separated by two interfaces.
 
\section{Concluding remarks}\lbl{sec.discuss}
\setcounter{equation}{0}

In this paper, we studied coupled Kuramoto oscillators on SW graphs. We showed that 
 in the limit as  the number of oscillators tends to infinity the Kuramoto equation
on SW graphs has a family of $q$-twisted state solutions, like those that have been studied
for the networks with $k$-NN coupling \cite{WilStr06, GirHas12}.
We used the continuum limit for the Kuramoto model on SW
graphs to study the stability of the $q$-twisted states. The linear stability analysis showed
that the randomization of the network connectivity promotes synchronization, while inhibiting
other $q$-twisted states. 

To study the role of the long-range connections in shaping  the spatial structure of solutions,
we developed a continuation method, which reveals the transformation of $q$-twisted 
state solutions as the randomization parameter is varied. The continuation showed that
adding random long-range connections results in spatial patterns composed of plateaus
separated by interfaces. The appearance of plateaus is another manifestation of 
the synchronizing effect of the long range connections. The solutions evolving from the $q$-twisted
states with $q\neq 0$ can not become completely synchronous, because their nonzero degree
is preserved under continuous deformation. This leads to the formation of interfaces.
This scenario is different from known mechanisms of interface formation in scalar
reaction-diffusion systems, where the sharpness of the interface is controlled by a
small parameter \cite{Fife-layers}, nor does it involve higher order
differential operators as in the Cahn-Hilliard equation \cite{CahnHill}.
Instead, the present mechanism is based on the properties of coupled
systems with nonlocal interactions \cite{Med13}. The kernel (\ref{GM}) for
the continuum limit (\ref{cKur}) corresponding to the SW networks is a piecewise
continuous function on $I^2$.  Theorem~3.2 in \cite{Med13} does not 
exclude discontinuous in $x$ solutions of (\ref{cKur}). In particular, the sequences shown 
in Figs. \ref{f.2} and \ref{f.3} may be approaching such discontinuous solutions, and 
may develop sharp interfaces in the process.
As a related observation, we note that spatial patterns shown in Figs. \ref{f.2} and 
\ref{f.3} combine regions of synchronous 
and asynchronous behaviors when viewed in the original coordinates
(see (\ref{preKuramoto}) with $\omega_i=\omega\neq 0, i\in [n]$). 
In this respect, they are 
similar to the chimera states \cite{KurBat02}. Therefore, the results of this work elucidate pattern-formation
mechanisms in nonlocally coupled dynamical systems.  

\vskip 0.2cm
\noindent
{\bf Acknowledgements.} 
This work was supported in part by the NSF grant  DMS 1109367.
 
\vfill\newpage
\bibliographystyle{amsplain}
\providecommand{\bysame}{\leavevmode\hbox to3em{\hrulefill}\thinspace}
\providecommand{\MR}{\relax\ifhmode\unskip\space\fi MR }
\providecommand{\MRhref}[2]{%
  \href{http://www.ams.org/mathscinet-getitem?mr=#1}{#2}
}
\providecommand{\href}[2]{#2}

\end{document}